%% file: main.tex
\documentclass[11pt]{article}

\usepackage[a4paper,margin=1in]{geometry}
\usepackage{amsmath,amssymb,amsfonts,amsbsy,latexsym,amsthm}
\usepackage{newtxtext,newtxmath}
\usepackage{url}
\usepackage{makecell}
\usepackage{graphicx}
\graphicspath{{Fig/}}
\usepackage{epstopdf}
\usepackage{enumitem}
\usepackage[caption=false]{subfig}
\usepackage{tabularx}
\usepackage{booktabs}
\usepackage{xcolor}
\usepackage{algorithm}
\usepackage{algpseudocode}
\usepackage{float}
\usepackage{natbib}
\usepackage{authblk}
\usepackage{microtype}
\definecolor{linkblue}{RGB}{0,76,153}
\usepackage[
	colorlinks=true,
	citecolor=linkblue,
	urlcolor=linkblue,
	filecolor=linkblue,
	linkcolor=black
]{hyperref}

\title{SeisEvo: Evolution of Seismic Data Reconstruction Algorithms by Agents}

\author[1,2]{Yingjie Xu}
\author[1]{Siwei Yu\thanks{Corresponding author: siweiyu@hit.edu.cn}}
\author[1,3]{Jianwei Ma}

\affil[1]{School of Mathematics and Center of Geophysics, Harbin Institute of Technology, Harbin, Heilongjiang, China}

\affil[2]{Department of Mathematics, National University of Singapore, Singapore}

 \affil[3]{School of Earth and Space Sciences, Peking University, Beijing, China}

\date{}

\begin{document}

\maketitle

\begin{abstract}
Classical seismic data reconstruction relies on manually designed structural
priors and iterative operators, whose coupled design space is far larger than
manual trial and error can explore systematically. Deep-learning methods encode the reconstruction rules in learned weights rather than in an explicit operator that can be inspected and modified. We propose SeisEvo (Seismic Algorithm Evolution), which does not optimize
a single reconstruction result but searches for the algorithm that produces it.
Starting from a classical reconstruction algorithm, an LLM-driven multi-agent search modifies only the components that the user has opened for editing, without prescribing the mechanism to be discovered. Candidates that violate the physical constraints of the task are rejected outright, and the remaining ones are scored by execution. The output is
neither an agent system nor a neural network, but a standalone white-box
algorithm that requires no agent or neural network at inference time. For
interpolation without added noise, the search discovered a
residual-gated, phase-aligned dip-consistency projection; Evo-POCS improves the
SNR over classic POCS by 3.49 dB on average across missing ratios from 30\% to
70\%. For simultaneous interpolation and denoising, it discovered a
reliability-grouped singular-value shrinkage; Evo-MSSA improves the average
reconstruction SNR by more than 7 dB over classic MSSA and by more than 3 dB
over a stronger rank-reduction baseline. Both operators retain their gains on
data not used during the search. To the best of our knowledge, this is the
first study to formulate the design of a seismic reconstruction operator as a
constrained, LLM-driven program evolution task. Agentic algorithm evolution can
thus complement deep learning in discovering explicit, inspectable, and
deployable seismic processing algorithms.
\end{abstract}


\input{sections/01_introduction}

\input{sections/02_method}
\input{sections/03_experiments}
\input{sections/03_experiments2}
\input{sections/04_discussion_conclusion}
\input{sections/05_acknowledgments}
\input{sections/appendix}

\bibliographystyle{plainnat}
\bibliography{bibliography}

\end{document}

%% file: sections/01_introduction.tex
\section{Introduction}

Seismic data reconstruction is an important processing step for improving the
spatial sampling quality of seismic records and for restoring the continuity of
subsurface reflection events. Because of surface obstacles, economic cost, and
practical acquisition conditions, field seismic data commonly suffer from
missing traces and irregular spatial sampling, which degrade the reliability of
subsequent imaging, inversion, and interpretation \citep{herrmann2008non,trad2009five}. Since the missing
samples are not observed, seismic data reconstruction is an underdetermined
inverse problem, and prior information about seismic structure must be
introduced to constrain the solution, such as transform-domain sparsity
\citep{abma20063d,herrmann2008non}, local coherence \citep{spitz1991seismic,naghizadeh2007multistep}, and structured low-rank behavior \citep{oropeza2011simultaneous}. Although data-driven methods have advanced rapidly in recent years, many
seismic reconstruction methods, as in many other seismic processing steps,
still rely on explicitly designed mathematical operators and structural
priors: how a prior is represented, through which computational operator it
is imposed, and how the estimate is kept consistent with the observations are
usually decided manually from domain experience. Reconstruction performance
therefore depends largely on how effectively these operators encode seismic
structure.

Classical reconstruction methods encode these structural priors explicitly as
interpretable algorithmic procedures. Based on transform-domain sparsity, the
projection onto convex sets (POCS) method \citep{abma20063d} recovers missing
traces by alternating transform-domain thresholding with data-consistency
enforcement. Based on local coherence, prediction-filter and local-slope-based methods
\citep{spitz1991seismic,naghizadeh2007multistep,xu2024dealiased} instead extrapolate missing
traces directly from the predictable relation between neighboring traces.
Based on structured low-rank behavior, multichannel singular spectrum analysis
(MSSA) \citep{oropeza2011simultaneous} applies a truncated low-rank projection
to block-Hankel embeddings, restoring the low-rank structure formed by
coherent reflection events and suppressing the rank increase caused by missing
traces and noise. These methods rest on explicit structural assumptions,
follow transparent computational steps, and are easy to deploy. POCS with
exponential hard thresholding \citep{gao2010irregular} and block-Hankel MSSA
with truncated singular value decomposition \citep{oropeza2011simultaneous}
are widely used as reference implementations of the two families, and are
adopted as the seed algorithms in this work.

Building on these classical templates, many manually designed improvements
have been proposed. For POCS, they mainly concern the choice of the
sparsifying transform \citep{wang2014dreamlet,zhang20153d}, the form of the
thresholding or shrinkage function, and the threshold decay schedule used
during the iterations \citep{gao2010irregular,gao2013convergence}. For MSSA,
they mainly concern the embedding and local window design, the rank selection,
and the singular-value processing rule
\citep{zhang2016multi,wu2018adaptive}, leading further to enhanced low-rank
methods such as damped rank reduction \citep{huang2016damped,chen2023drr} and
optimally damped rank reduction (ODRR) \citep{chen2020odrr}. These improvements are
introduced through expert-guided, mechanism-specific design and are usually
validated under specific acquisition conditions or noise levels. However, most
studies address only one or a few dimensions of the design space, and their
effectiveness depends on the particular data distribution, sampling density,
and signal-to-noise ratio, so they may require retuning or redesign under
different conditions. Moreover, these design factors interact: the threshold
form and the decay schedule, or the rank selection and the singular-value
shrinkage, jointly determine performance, so adjusting each factor in
isolation is not sufficient for exploring their combined effects. The design
space left open by these classical templates is therefore far larger than
manual exploration can systematically cover.

In recent years, deep learning has provided an alternative route for seismic
data reconstruction. By learning data-driven reconstruction mappings or priors,
such methods have been applied to seismic data interpolation and noise
suppression \citep{yu2019deep,wang2019deep,kaur2021seismic,cheng2025multitask}. To reduce the reliance on paired complete data,
unsupervised and self-supervised methods further construct training signals
from unlabeled observations or from the data to be reconstructed themselves
\citep{liu2021deep,meng2022self,abedi2022multidirectional,chen2024combining,xu2026unsupervised}. Seismic foundation models have also emerged, which learn general
seismic representations through large-scale self-supervised pretraining and
transfer them to different downstream tasks such as interpolation, denoising,
and inversion \citep{sheng2025seismic,cheng2025generative}.

These methods differ in nature from classical reconstruction algorithms. The
performance of supervised methods depends on the data distribution covered by
the training set, and their generalization needs to be re-examined when the
acquisition conditions, noise levels, or geological structures differ
appreciably from the training data. Unsupervised and self-supervised methods
reduce the reliance on fully labeled data, but their performance is still
affected by the network architecture, the regularization, and the training
strategy. For most end-to-end methods, the main reconstruction rules are
distributed across the network weights and the training procedure, and are
therefore difficult to write down, inspect, and modify under given constraints
in the way classical operators are. Deep learning has thus not removed manual design; it has moved part of the design effort from the reconstruction operator to the network architecture, the loss function, and the training strategy. Classical and deep-learning methods still share a common feature: both optimize within a computational structure that is defined in advance by humans. In most existing studies on seismic reconstruction, the task-relevant algorithmic structure is still designed manually, and treating the algorithmic structure itself as the object of search remains rare.

In recent years, large language model (LLM)-driven program search and evolution
have opened a new possibility for algorithm design. These methods represent a
candidate algorithm as executable code, let an LLM propose modifications, and
then run and score each candidate with an automated evaluator, using the
resulting feedback to guide the subsequent search. FunSearch showed that this
paradigm can discover improved mathematical constructions and heuristics in
combinatorial mathematics and combinatorial optimization
\citep{romeraparedes2024mathematical}. EoH and ReEvo subsequently improved
heuristic generation, the use of feedback, and search efficiency, although the
object of evolution there is typically a heuristic rule inside a fixed
algorithmic pipeline \citep{liu2024evolution,ye2024reevo}. In an applied
setting, Eureka evolved reward code and obtained reward functions that
outperform human-engineered ones on most robotic control tasks
\citep{ma2024eureka}. AlphaEvolve further extended the paradigm to entire code
bases and more general algorithm optimization tasks, and applied it to problems
such as matrix multiplication, data center scheduling, and hardware design
\citep{novikov2025alphaevolve}. Unlike end-to-end neural approaches, such a search returns a program that can be executed and inspected on its own rather than a set of learned network weights; interpretability, however, is not obtained automatically, but depends on how the program is represented and on the constraints imposed during the search.

These advances indicate that the algorithm itself can become the object of
search. Seismic data reconstruction offers a well-suited setting for this
paradigm. First, reconstruction quality can be measured directly by metrics such as SNR and SSIM on the prescribed search-time blocks for which complete references are available, so that each candidate can be executed and assigned an explicit score. Second, classical reconstruction algorithms have a clear and modular structure, so that mechanisms such as the thresholding rule, the decay schedule, and the low-rank
projection can be designated individually as editable components. Third,
seismic reconstruction carries admissibility conditions that are explicit and
decidable: how the observations are used, whether the sampling mask is left
unchanged, and whether the output is real-valued are all determined once the
task is given, and can therefore be checked programmatically.

Applying this paradigm to seismic reconstruction, however, is not a matter of
simply running a generic search. When domain legality conditions are absent, an
increase in score does not necessarily correspond to an improvement of the
algorithm: a candidate may reach a physically meaningless score by corrupting
the observations or altering the sampling mask; it may raise a metric through
unexplained free constants or dataset-specific shortcuts rather than by finding
a new mechanism. For seismic reconstruction, legality depends not only on
mathematical and numerical correctness but also on physical measurement
conditions, so how these conditions are formalized and enforced as part of the
search is itself part of the problem. To date, the use of LLM-driven program
evolution to systematically discover standalone, white-box seismic
reconstruction operators remains largely unexplored, and validation across
different mathematical mechanisms and different reconstruction task settings is
especially lacking.

To address these issues, we propose SeisEvo (Seismic Algorithm Evolution), a
domain-grounded framework for the autonomous discovery of seismic
reconstruction algorithms. SeisEvo takes the reconstruction algorithm itself,
rather than a particular reconstructed result, as the object of search: instead
of optimizing the output of a single reconstruction, it searches for an
algorithm that produces high-quality reconstructions. A run is specified by a
domain expert through a transparent classical seed algorithm $A_0$, an editable
operator surface $\mathcal{S}$, a reconstruction-quality objective $O$, a
constraint set $\mathcal{C}$, and a search budget $T$. Here $\mathcal{S}$
states which mechanisms may be modified, such as the thresholding rule, the
decay schedule, and the low-rank projection, whereas the task-specific
data-fidelity requirements remain constrained throughout the search. The expert thus
does not specify the final form of the operator in advance, but defines the
search boundary, the physical contracts, and the scoring rules, leaving the
mechanism itself to be found by the search.

SeisEvo uses the Evolutionary Ensemble of Agents (EvE)
\citep{yu2026evolutionary}, a decentralized evolutionary framework that
organizes multiple coding agents to iteratively propose and refine candidate
programs. EvE and EvE-based workflows have been used for scientific problems
including fluid-control design, operator learning, and PDE symbolic discovery \citep{sun2026self,yang2026harness,yu2026agentic}. Here it serves as the
generic multi-agent search backend. Our contribution lies in the domain-specific framework for seismic algorithm
evolution, including the algorithm representation, the editable operator
surface, the legality constraints, the execution-based scoring, and the
external validation protocol. When the search ends, the
highest-scoring legal candidate is returned as $A^\star$, a standalone
white-box reconstruction operator. Its computational steps and introduced
parameters can be written down, inspected, and reproduced, and it requires no
agent, prompt, or neural network at inference time. We instantiate SeisEvo on two classical algorithm families with different
mathematical foundations: POCS based on Fourier-domain sparsity and MSSA based
on block-Hankel low-rank structure. The resulting operators are referred to as
Evo-POCS and Evo-MSSA, respectively.

This paper makes three main contributions. First, we propose SeisEvo, an
algorithm evolution framework for seismic data reconstruction that formulates
the design of a reconstruction algorithm as a domain-constrained autonomous
discovery task. Starting from a classical seed algorithm, SeisEvo searches for
new operator mechanisms that improve reconstruction quality, while
task-specific data-fidelity requirements remain enforced throughout the search. Second, we establish an executable search and validation protocol that uses
hard legality gates and requires mechanism-level changes with traceable
parameter origins. External data are accessed only after the search, once the
selected operator has been fixed. Third, we validate SeisEvo on two classical algorithm families with different
mathematical foundations. For interpolation of a field dataset without added
synthetic noise, Evo-POCS improves the SNR over classic POCS by 3.49 dB on
average across missing ratios from 30\% to 70\%. For simultaneous interpolation and denoising,
Evo-MSSA improves the mean reconstruction SNR by more than 7 dB over
classic MSSA and by more than 3 dB over the ODRR baseline over the tested noise range.
Both evolved operators retain their performance gains on synthetic
and field data that were not used during the search. To the best of our
knowledge, this is the first reported application of LLM-driven constrained
program evolution to the discovery of standalone seismic reconstruction
operators.

The remainder of this paper is organized as follows. Section 2 introduces the
seismic reconstruction problem and the SeisEvo framework, including the task
specification and constraint design. Section 3 presents the POCS and MSSA case
studies and their experimental results. Section 4 discusses what the search contributed, the scope and cost of the
protocol, and the open problems it leaves. Section 5 concludes the paper.

%% file: sections/02_method.tex
\section{Method}
\label{sec:method}

\subsection{Seismic data reconstruction problem}
\label{sec:method_problem}
Let $X$ denote a complete seismic volume of size
$n_t\times n_x\times n_y$, and let $M$ be a binary sampling mask of the same
size, where $M(t,x,y)=1$ indicates an acquired sample and $M(t,x,y)=0$ a
missing one. Since traces are missing as whole columns, $M$ is constant along
the time axis. Let $N$ denote additive noise on the acquired samples. The
observed incomplete volume is
\begin{equation}
	Y = M \odot (X+N) ,
	\label{eq:observation}
\end{equation}
where $\odot$ denotes element-wise multiplication, and $N=0$ gives the
noise-free interpolation setting. The task is to estimate a complete volume
$\widehat X$ from $(Y,M)$. Because the missing traces are unobserved, the
problem is underdetermined, and it is commonly formulated as a regularized
inverse problem,
\begin{equation}
	\widehat X
	=
	\arg\min_{Z}\;
	\frac{1}{2}\,\bigl\| Y - M \odot Z \bigr\|_F^2
	\;+\;
	\lambda\, R(Z) ,
	\label{eq:inverse_problem}
\end{equation}
where the first term enforces fidelity to the acquired traces and $R(Z)$ is a
regularization term encoding prior knowledge about the structure of seismic
data, balanced by $\lambda$. Two representative and widely used priors are transform-domain sparsity and
structured low rank: the former assumes that seismic data admit compact
representations in a suitable transform domain, while the latter assumes that
coherent seismic events give rise to approximately low-rank structures after
suitable embedding, which missing traces and noise tend to break. 

Problem~\eqref{eq:inverse_problem} motivates a broad class of iterative
schemes that alternate between enforcing the assumed prior and honoring the
observations,
\begin{equation}
	Z^{(j+1)}
	=
	\mathcal P_{\mathrm{data}}
	\!\left(
	\mathcal T_{\mathrm{prior}}^{(j)}(Z^{(j)})
	\right),
	\label{eq:two_operator_template}
\end{equation}
where $\mathcal T_{\mathrm{prior}}^{(j)}$ is an iteration-dependent operator
enforcing the seismic prior, and $\mathcal P_{\mathrm{data}}$ is the
task-specified data-fidelity step. In noise-free interpolation it is the hard
projection
\begin{equation}
	\mathcal P_{\mathrm{data}}(U) = Y + (1-M)\odot U ,
	\label{eq:data_projection}
\end{equation}
which restores the acquired traces exactly; when the acquired samples are
noisy, the hard projection is replaced by a task-specific relaxed fidelity
step that balances the observations and the current prediction. Its concrete
form is given in the corresponding case study. POCS and MSSA are two
representative classical solvers associated with these priors and serve as the
seed algorithms in this work: POCS realizes
$\mathcal T_{\mathrm{prior}}^{(j)}$ through transform-domain thresholding,
encoding the sparsity prior, while MSSA realizes it through a truncated low-rank projection of block-Hankel
embeddings, encoding the low-rank prior.

Taking these two algorithms as templates, numerous studies have developed
further improvements. Improvements to POCS mostly focus on the sparsifying
transform, the thresholding or shrinkage function, and the decay schedule
\citep{gao2010irregular, wang2014dreamlet,zhang20153d,dong2025robust}, with hard thresholding and
exponential decay as the standard baseline. Improvements to MSSA mostly focus on the rank-reduction operator and rank selection
\citep{huang2016damped,zhang2016multi,bayati20233}, with truncated SVD and a
fixed rank as the standard baseline. Such designs are interpretable and often effective, but each modification explores only a small portion of the possible design choices: tuning relies heavily on expert experience, and a setting that works under one sampling or noise condition rarely transfers unchanged to another. The design space left open by these templates is far larger than what
manual exploration can cover.

The recent emergence of LLM-based coding agents makes it feasible to explore
this design space automatically. Motivated by this, we propose SeisEvo, which
turns the open design space into a search space while keeping the
mask-based structure of the task-specific data-fidelity step and the
physical-legality constraints fixed. We instantiate it on the two seed algorithms introduced above: POCS with the
sparsity prior and MSSA with the low-rank prior. These two solvers are
mechanistically distinct, which lets us verify that the framework operates
beyond a single algorithm family. The framework is detailed next.

\subsection{The SeisEvo framework}
\label{sec:method_seisevo}
We introduce \textbf{SeisEvo} (Seismic Algorithm Evolution), a framework for
the autonomous discovery of interpretable seismic reconstruction algorithms.
SeisEvo treats the reconstruction algorithm itself, rather than a particular
reconstructed result, as the object of search. Given a transparent classical
seed algorithm $A_0$ of the form~\eqref{eq:two_operator_template}, a SeisEvo
run with round budget $T$ is written as
\begin{equation}
	A^\star = \operatorname{SeisEvo}\left(A_0,\;\mathcal S,\;O,\;\mathcal C;\;T\right),
	\label{eq:seisevo_formulation}
\end{equation}
where $\mathcal S$ denotes the editable parts of the algorithm, such as the
thresholding function, the decay schedule, or the low-rank projection, while
the mask-based structure of the task-specific data-fidelity step remains fixed.
$O$ is the objective, a measured reconstruction-quality score, and
$\mathcal C$ is the constraint set, covering physical-legality
constraints, parameter-provenance rules, anti-leakage requirements, and the
search-time scoring protocol. The budget $T$ is the number of search rounds.
The returned $A^\star$ is the highest-scoring legal candidate, namely the
evolved white-box algorithm.

The defining principle is that the task specification states what counts as a
legal and worthwhile algorithm, not what the algorithm should be. The
objective $O$ and constraint set $\mathcal C$ determine only how
candidates are admitted and scored. They contain no candidate code, no target
operator, and no fixed implementation template, so the mechanism itself is
left for the search to discover. 



The output is a standalone white-box algorithm whose computational steps and
introduced parameters can be inspected and executed without agents, prompts,
or neural networks at inference time.

\begin{figure}[H]
	\centering
	\includegraphics[width=1\textwidth]{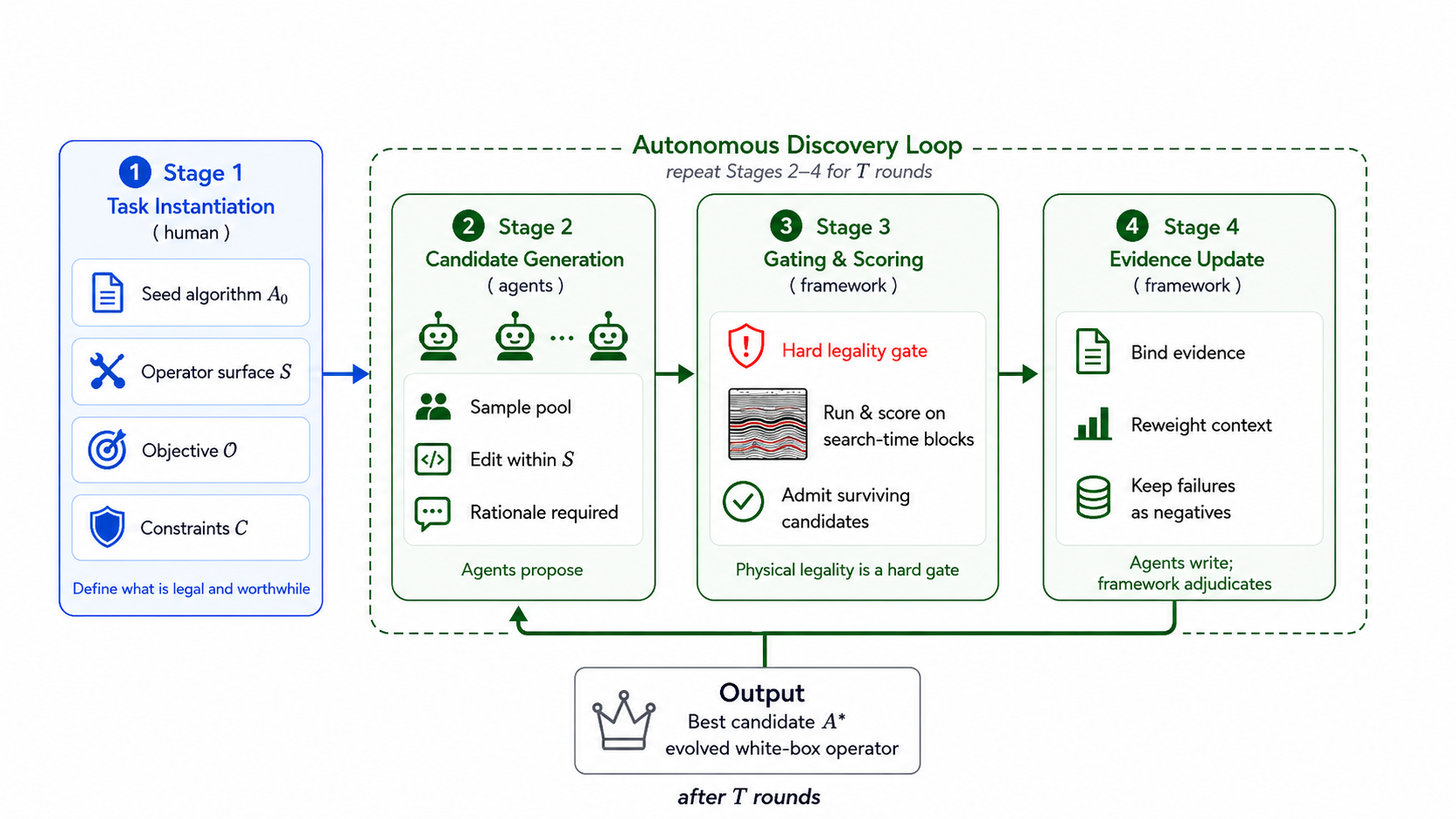}
	\caption{The SeisEvo framework. A one-time task instantiation (seed
		algorithm and specification) is followed by a closed loop of candidate
		generation, gating and scoring, and evidence binding with context update,
		repeated for $T$ rounds. The best candidate is returned as a standalone
		white-box operator. External validation, including full-volume generalization
		and robustness tests under different sampling and noise conditions, is
		performed once after the search and is not part of the loop.}
	\label{fig:seisevo_framework}
\end{figure}

A SeisEvo run consists of four stages: \emph{task instantiation},
\emph{candidate generation}, \emph{gating and scoring}, and \emph{evidence
	binding and context update} (Fig.~\ref{fig:seisevo_framework}). Stage~1 is
performed once by a human. Stages~2 to 4 form a closed loop that repeats for
$T$ rounds. The four stages answer four questions in order: what algorithm is
worth searching for, what to try next, whether a candidate qualifies to be
compared, and what a completed trial counts as evidence for. External
validation is not part of the loop; it is performed once on the final
algorithm after the search ends.

\textbf{Stage 1: Task instantiation (human).} A human defines what counts as a
legal and worthwhile algorithm, not what the algorithm should be. This
includes the seed algorithm $A_0$, the editable components $\mathcal S$, the
objective $O$, and the constraint set $\mathcal C$ defined above.

\textbf{Stage 2: Candidate generation (agents).} Multiple coding agents work
in parallel. Each agent reads sampled high-scoring candidates, failed
candidates, diverse candidates, and the current search context. It then
proposes new candidates by editing the algorithm only within $\mathcal S$.
Agents may also propose updates to the search context for later rounds, but
these are treated as prospective hypotheses rather than adjudicated evidence.
To support attribution and inspection, every candidate carries a mandatory
rationale: what mechanism it changes, where each new parameter comes from, and
why the change may help.

\textbf{Stage 3: Gating and scoring (framework).} Physical legality is a hard
gate, not a soft penalty. Candidates are rejected outright if they violate the
task-specified data-consistency requirements, for example by breaking observed
traces or altering the sampling mask in interpolation, or if they fail to run
or diverge numerically. Surviving candidates are run on the search-time
sub-blocks, scored by a common metric, and then admitted to the candidate pool.
Rejected and failed candidates are kept as negative examples.

\textbf{Stage 4: Evidence binding and context update (framework).} The
framework records each candidate's rationale, score, or failure log. It then
updates which candidates and mechanism rationales are more likely to be shown
to agents in the next round. This stage produces no new algorithmic hypotheses
and contains no language-model judgment; it applies fixed programmatic rules to
execution-derived scores and failure records. Agents only propose; all
adjudication is done by the framework from real execution.

After $T$ rounds, the highest-scoring legal candidate under $O$ is frozen as
the evolved algorithm $A^\star$, which then proceeds to external validation
outside the search.

\begin{algorithm}[H]
	\caption{SeisEvo design run}
	\label{alg:seisevo}
	\begin{algorithmic}[1]
		\Require seed algorithm $A_0$; editable surface $\mathcal S$;
		objective $O$; constraint set $\mathcal C$; rounds $T$
		\Ensure evolved white-box algorithm $A^\star$
		\State \textbf{Stage 1 (human):} instantiate $(A_0,\mathcal S,O,\mathcal C)$
		\State pool $\gets \{A_0\}$
		\For{$t = 1,\ldots,T$}
		\State \textbf{Stage 2 (agents):} agents read sampled candidates,
		failure records, and context, then propose new candidates by editing only
		within $\mathcal S$, each with a rationale and parameter provenance
		\State \textbf{Stage 3 (framework):} reject candidates that violate
		$\mathcal C$, fail to run, or diverge; otherwise run them on search-time
		sub-blocks, score them by $O$, and add them to the pool
		\State \textbf{Stage 4 (framework):} record each candidate's rationale
		and outcome; update next-round context and read priorities by fixed
		programmatic rules
		\EndFor
		\State \Return the best-scoring candidate $A^\star$ under $O$
	\end{algorithmic}
\end{algorithm}

\subsection{Task specification}
\label{sec:method_spec}
Besides the seed algorithm $A_0$ and the editable surface $\mathcal S$, a
SeisEvo run is specified by the objective $O$ and the constraint set
$\mathcal C$. Together they state what counts as a valid and worthwhile
algorithm, without prescribing the mechanism to be found.

\textbf{Objective $O$.} A scalar reconstruction-quality score. It may be a
single metric, such as the signal-to-noise ratio (SNR), or a weighted
combination of several metrics, such as SNR together with SSIM or MSE. A
composite scalar objective may guide the search toward more robust results
than a single metric alone \citep{novikov2025alphaevolve}. The
concrete scoring design is task-dependent and is given in the experiments.

\textbf{Physical legality.} For the interpolation tasks considered here,
observed traces must be preserved exactly; for noisy reconstruction tasks,
this hard projection can be replaced by a noise-aware fidelity rule. In
addition, every candidate must leave the sampling mask unchanged, return a
real-valued volume, and keep the fixed input--output interface. These are hard
gates: violating any one rejects the candidate outright. Dataset-specific
shortcuts and any use of test data during the search are likewise forbidden.

\textbf{Interpretability rules.} Two rules keep the discovered algorithm
auditable. First, \emph{no pure parameter tuning}: the specification prohibits
candidates whose only claimed change is retuning existing scalar constants,
for example making a threshold decay faster. Changing the functional form of a
threshold, schedule, or projection counts as a mechanism-level modification.
Second, \emph{parameter provenance}: every new free parameter must be inherited
from the seed protocol, derived from observed statistics or mathematical
bounds, introduced solely for numerical stability, or selected from a
predeclared finite structural set. Unexplained free constants are forbidden.
The first rule prevents progress by tuning numbers; the second requires an
origin for any number that does change.

\textbf{Scoring during the search.} Candidates are scored only on search-time
sub-blocks extracted from the data; the external test volumes are never
touched until the search has ended.

These rules form the common basis shared by both case studies. The framework
does not fix the exact form of the constraints: for a different task, the
expert may add further physical or numerical constraints, such as preserving a
given frequency band or bounding amplitude changes, which act as hard gates in
the same way.

When the search ends, the highest-scoring legal candidate $A^\star$ is frozen
as the evolved algorithm: a standalone white-box solver that runs without any
agent, prompt, or neural network at inference time. Because candidates are
scored only on search-time sub-blocks during the search, we validate
$A^\star$ once after the search to examine whether its performance generalizes
beyond the data used to guide the search. In the experiments, this validation
combines full-volume generalization and robustness tests
under different sampling and noise conditions. We instantiate the protocol on POCS and
MSSA, yielding Evo-POCS and Evo-MSSA.

%% file: sections/03_experiments.tex
\section{Numerical experiments}

\subsection{Experimental protocol}

The data used for evolution are $64\times64\times64$ sub-blocks extracted from
complete 3D seismic volumes. During the search, candidate algorithms are
generated, scored, and used to update the context only on these sub-blocks.
After the search ends, the discovered algorithm is tested on seismic data
outside this set to assess its generalization.

\begin{figure}[H]
	\centering
	\includegraphics[width=1\textwidth]{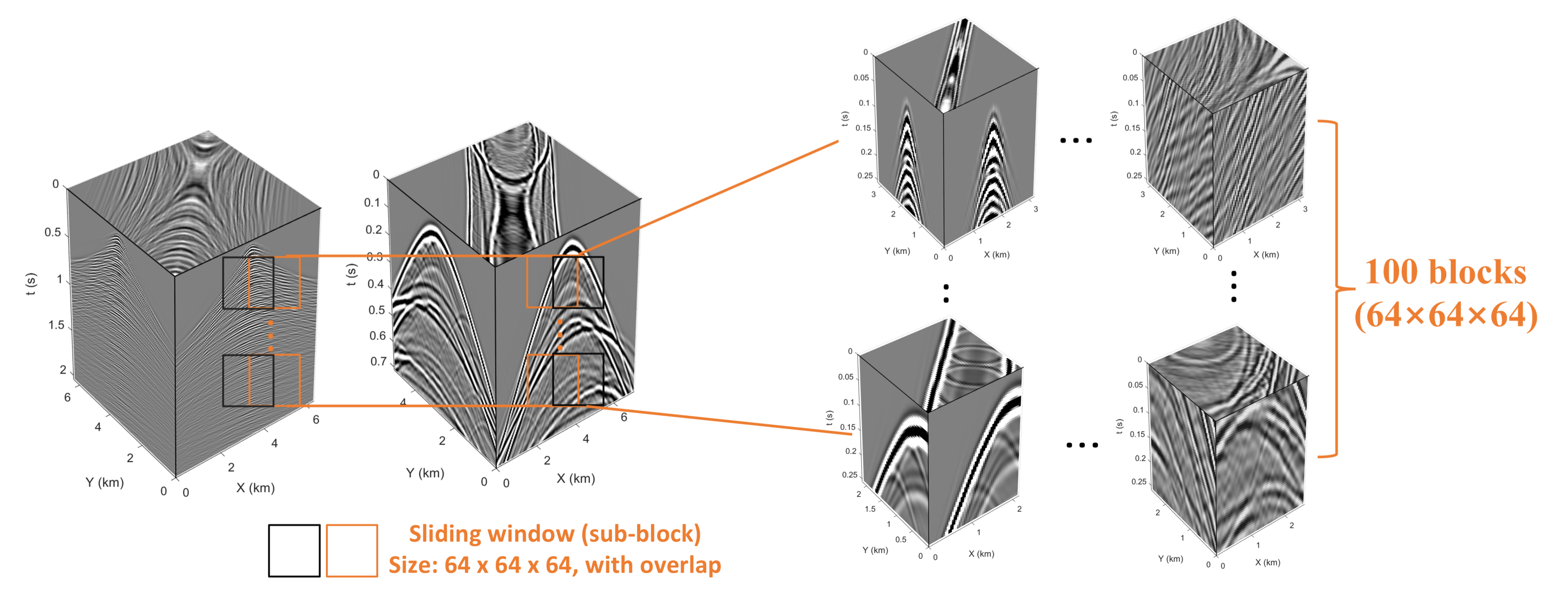}
	\caption{Search-time seismic sub-blocks extracted from complete 3D volumes.}
	\label{fig:seisevo_data}
\end{figure}

In seismic data reconstruction, the signal-to-noise ratio (SNR) is a common measure of reconstruction quality; we report SNR together with the structural similarity (SSIM) and the mean squared error (MSE):
\begin{equation}
	\mathrm{SNR} = 10\log_{10}
	\frac{\|X\|_F^2}{\|X - \widehat X\|_F^2},
	\label{eq:snr}
\end{equation}
\begin{equation}
\mathrm{MSE} = \frac{1}{N_{\mathrm{sam}}}
\|X - \widehat X\|_F^2,
	\label{eq:mse}
\end{equation}
\begin{equation}
	\mathrm{SSIM}(X, \widehat X) =
	\frac{(2\mu_X\mu_{\widehat X}+c_1)(2\sigma_{X\widehat X}+c_2)}
	{(\mu_X^2+\mu_{\widehat X}^2+c_1)(\sigma_X^2+\sigma_{\widehat X}^2+c_2)},
	\label{eq:ssim}
\end{equation}
where $X$ is the original complete data and $\widehat X$ the
reconstruction, $N_{\mathrm{sam}}$ is the total number of samples, and $\mu$
and $\sigma$ denote
means and (co)variances, and $c_1,c_2$ are small constants that avoid a zero
denominator. A higher SNR, a lower MSE, and an SSIM closer to one indicate
better reconstruction. For multiple volumes or cases, each metric is computed
per case and then averaged.

\subsection{SeisEvo test on POCS for interpolation}

\textbf{Task instantiation.} The first case study instantiates SeisEvo on the
POCS family for interpolation of randomly missing traces without adding
synthetic noise. In this setting, no additional noise term is introduced in
Eq.~\eqref{eq:observation}, i.e., $N=0$, so the observation reduces to
$Y = M \odot X$, where $M$ is a random
spatial sampling mask that drops whole traces. The four inputs of the method
are instantiated as follows. The seed $A_0$ is classic exponential
hard-threshold POCS. The editable surface $\mathcal S$ contains the
per-iteration thresholding and shrinkage rules, the decay schedule, and any
added projection or correction operator, but not the input--output interface or
the observed-trace restoration. The constraint set $\mathcal C$ requires the
observed traces to be restored exactly, the sampling mask to be preserved, and
the output to be real-valued. It also forbids test-set leakage and unexplained
free parameters. The objective is
\begin{equation}
	O_{\mathrm{POCS}} =
	0.7\,\operatorname{clip}\!\left(\frac{\mathrm{SNR}}{18},0,1\right)
	+ 0.3\,\mathrm{SSIM},
	\label{eq:pocs_objective}
\end{equation}
where 18~dB is an empirical reference used during the search to scale the SNR
term to a range comparable with SSIM.

\textbf{Classic POCS seed.}
The seed is classic POCS \citep{abma20063d}, with the exponential
threshold decay widely adopted in later work \citep{gao2010irregular}. After a
temporal Fourier transform, each frequency slice is reconstructed by alternating
a sparse projection in the spatial-frequency domain with an exact observed-trace
projection in the data domain. For a frequency slice $U_k^{(j)}$, one iteration
is
\begin{equation}
	V_k^{(j)}
	=
	\mathcal F^{-1}
	\left[
	H_{\tau_{k,j}}
	\left(
	\mathcal F\, U_k^{(j)}
	\right)
	\right],
	\label{eq:pocs_sparse_step}
\end{equation}
\begin{equation}
	U_k^{(j+1)}
	=
	Y_k + (1-M)\odot V_k^{(j)},
	\label{eq:pocs_data_step}
\end{equation}
where $\mathcal F$ is the 2D spatial Fourier transform over a frequency slice,
$H_{\tau}$ is hard thresholding, and $Y_k$ is the observed frequency slice. The
threshold $\tau_{k,j}$ decays exponentially from 90\% to 1\% of the maximum
transform-domain amplitude over 30 reconstruction iterations. The same
threshold schedule and iteration count are used by classic POCS and Evo-POCS.
Equation~\eqref{eq:pocs_data_step} is the data-consistency projection of
Eq.~\eqref{eq:data_projection} applied per frequency slice.

This seed exploits Fourier-domain sparsity but does not explicitly use other
properties of seismic data. Its reconstruction is therefore limited when
dipping or laterally continuous events are not well represented by Fourier
thresholding alone. The next part describes the addition that
SeisEvo discovers on top of this seed.

\begin{figure}[H]
	\centering
	\includegraphics[width=0.6\textwidth]{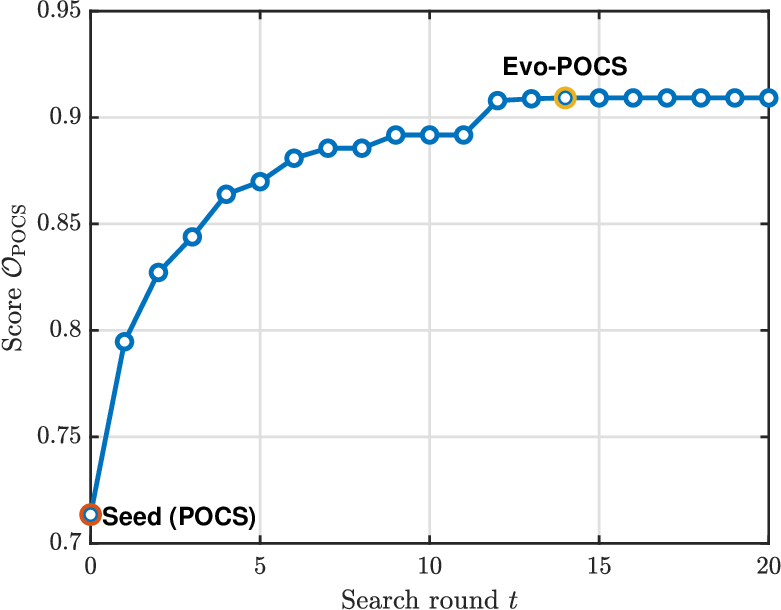}
	\caption{Search-time objective $O_{\mathrm{POCS}}$ over 20 search rounds.
		The score rises from 0.713 at the classic POCS seed to 0.909 at round 14 and
		remains unchanged afterwards.}
	\label{fig:pocs_score}
\end{figure}

\textbf{Search process.} Starting from the classic POCS seed, SeisEvo runs for
20 rounds. Figure~\ref{fig:pocs_score} shows the search-time objective
$O_{\mathrm{POCS}}$ as a function of the search round. The objective rises
quickly from 0.713 at the seed within the first ten rounds, reaches its best
value of 0.909 at round 14, and stays unchanged over the remaining six rounds,
indicating that the 20-round budget is sufficient for this task.
Correspondingly, the SNR on the training sub-blocks improves from 11.32 to
15.88~dB and SSIM from 0.911 to 0.972. This gain comes from a new structural
step introduced by the SeisEvo search, not from retuning the threshold or decay
parameters.

\textbf{Evo-POCS.} The new structural step introduced by the
search is a data-driven local dip-consistency correction, applied after each
Fourier thresholding step. In one sentence, Evo-POCS is classic POCS
constrained by local dip consistency: it moves a missing sample toward a local
wavefield prediction only where one dip direction is clearly more
self-consistent than the others.

Algorithm~\ref{alg:evopocs} summarizes the deployed Evo-POCS at the
frequency-slice level. In each iteration, after the classic POCS sparse
projection, a local phase relation is estimated from observed neighboring
traces for a fixed set of candidate dip directions; each direction predicts the
missing sample by phase-aligned continuation; the directions whose predictions
best match the observed samples are combined into a consensus prediction; and a
confidence weight, set by how clearly the best direction outperforms the rest,
controls how far the estimate moves toward this prediction. The observed traces
are restored exactly at each step. Implementation details of the phase
coefficient, residual weighting, and fallback rules are given in Appendix
~\ref{sec:appendix_evopocs}.

\begin{algorithm}[H]
	\caption{Evo-POCS reconstruction for one frequency slice}
	\label{alg:evopocs}
	\begin{algorithmic}[1]
		\Require observed slice $Y_k$, mask $M$, iterations $J$, direction bank $\mathcal D$
		\Ensure reconstructed slice $U_k$
		\State estimate local phase coefficients $c_d(u)$ for each direction
		$d\in\mathcal D$ from observed trace pairs
		\Comment{depends only on $Y_k$ and $M$}
		\State $U_k^{(0)} \gets Y_k$
		\For{$j = 0,\ldots,J-1$}
		\State $V \gets$ Fourier hard-threshold sparse projection of $U_k^{(j)}$
		\Comment{classic POCS step}
		\For{each direction $d \in \mathcal D$}
		\State predict $P_d$ from $V$ by phase-aligned continuation
		\State measure residual $R_d$ on observed samples
		\EndFor
		\State form consensus $P_\star$ from selected lowest-residual directions
		\State set confidence map $\kappa$ from residual contrast
		\State $V \gets V + \kappa \odot (P_\star - V)$
		\Comment{dip-consistency correction}
		\State $U_k^{(j+1)} \gets Y_k + (1-M)\odot V$
		\Comment{restore observed traces}
		\EndFor
		\State \Return $U_k^{(J)}$
	\end{algorithmic}
\end{algorithm}

After all frequency slices are reconstructed, an inverse temporal Fourier
transform gives the real-valued volume. The correction is not an explicit
isotropic smoothing operator, nor does it retune the threshold; instead, it uses the directional
continuity present in the data itself, deciding along which direction and by how
much to correct each missing sample directly from the observed data. Every part
of the operator has a clear seismic meaning: the Fourier threshold corresponds
to the sparsity assumption, the observed-trace projection keeps the measured
data unchanged, the candidate direction set represents the local dip of seismic
events, the phase estimate compensates the phase difference between neighboring
traces, and the residual and confidence ensure that the dip prior is trusted
only along the directions the observed data support. Unlike existing methods
that rely on dip continuity, the operator in Evo-POCS was discovered
autonomously by SeisEvo: without any prescribed mechanism, the search
constructed the full procedure of directional estimation, consensus, and gating
directly from the observed data, and integrated it into the POCS framework with
a clear improvement.

\begin{figure}[H]
	\centering
	\includegraphics[width=1\textwidth]{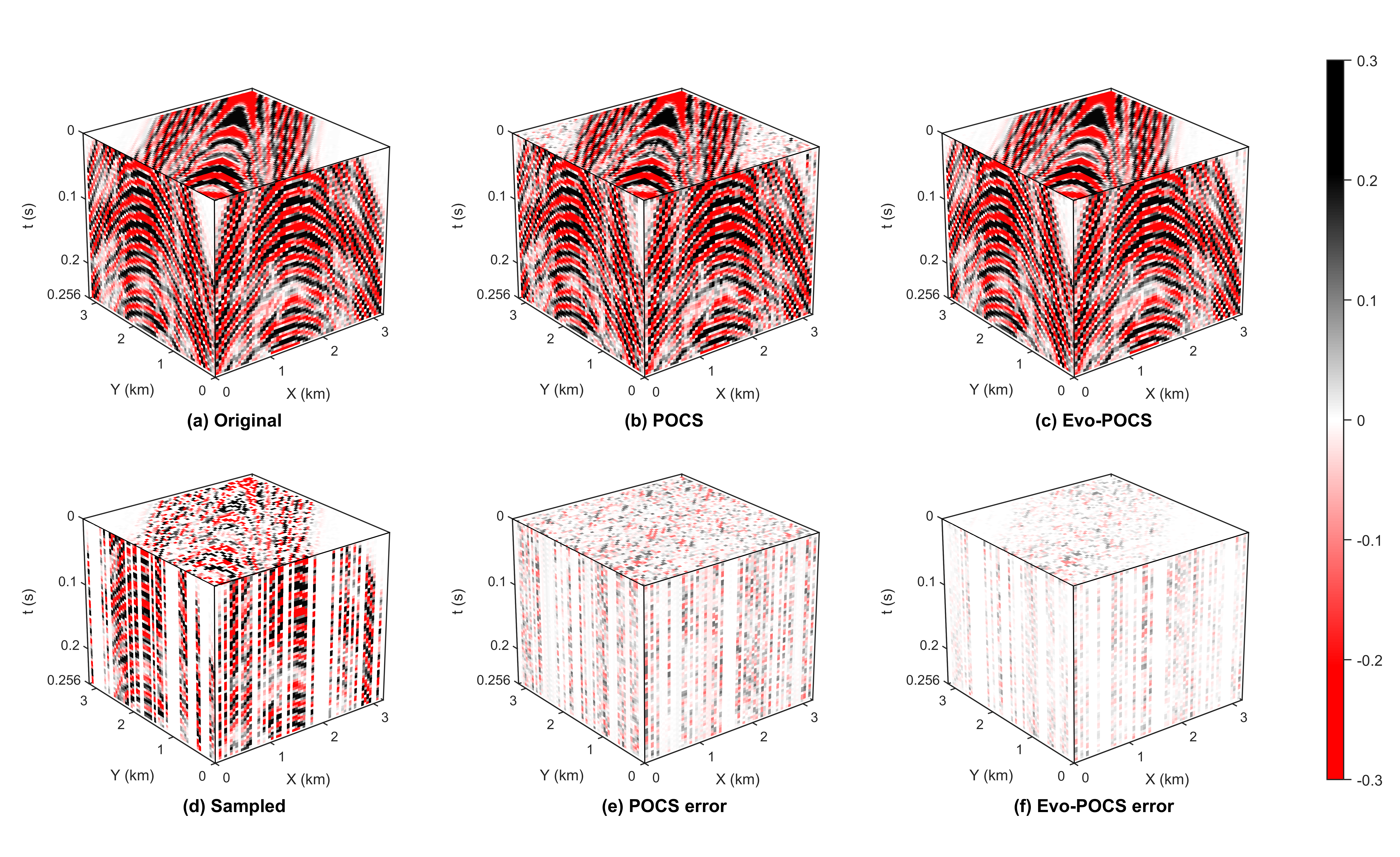}
	\caption{Reconstruction comparison on a $64\times64\times64$ synthetic
		sub-block under 50\% random sampling. (a) Original complete data;
		(b) POCS reconstruction; (c) Evo-POCS reconstruction; (d) 50\% randomly
		sampled data; (e) POCS error; (f) Evo-POCS error.}
	\label{fig:syn_pocs_compare}
\end{figure}

\textbf{Reconstruction results.} Figure~\ref{fig:syn_pocs_compare} compares
classic POCS and Evo-POCS on a $64\times64\times64$ sub-block under 50\% random
sampling. Panel (a) is the original complete data, (d) the 50\% randomly sampled
data, and (b) and (c) the reconstructions by classic POCS and Evo-POCS,
respectively. Both methods reconstruct the data effectively, but Evo-POCS
clearly yields better continuity of the local events, and is notably smoother on
the time slices. The error panels (e) and (f) further highlight this difference:
Evo-POCS produces less leakage, a smaller reconstruction error, and lower
residual noise. Quantitatively, the SNR on this sub-block improves from
11.78~dB for classic POCS to 18.26~dB, SSIM from 0.895 to 0.986, and MSE
decreases from $2.3\times10^{-3}$ to $5.0\times10^{-4}$.

\begin{figure}[H]
	\centering
	\includegraphics[width=1\textwidth]{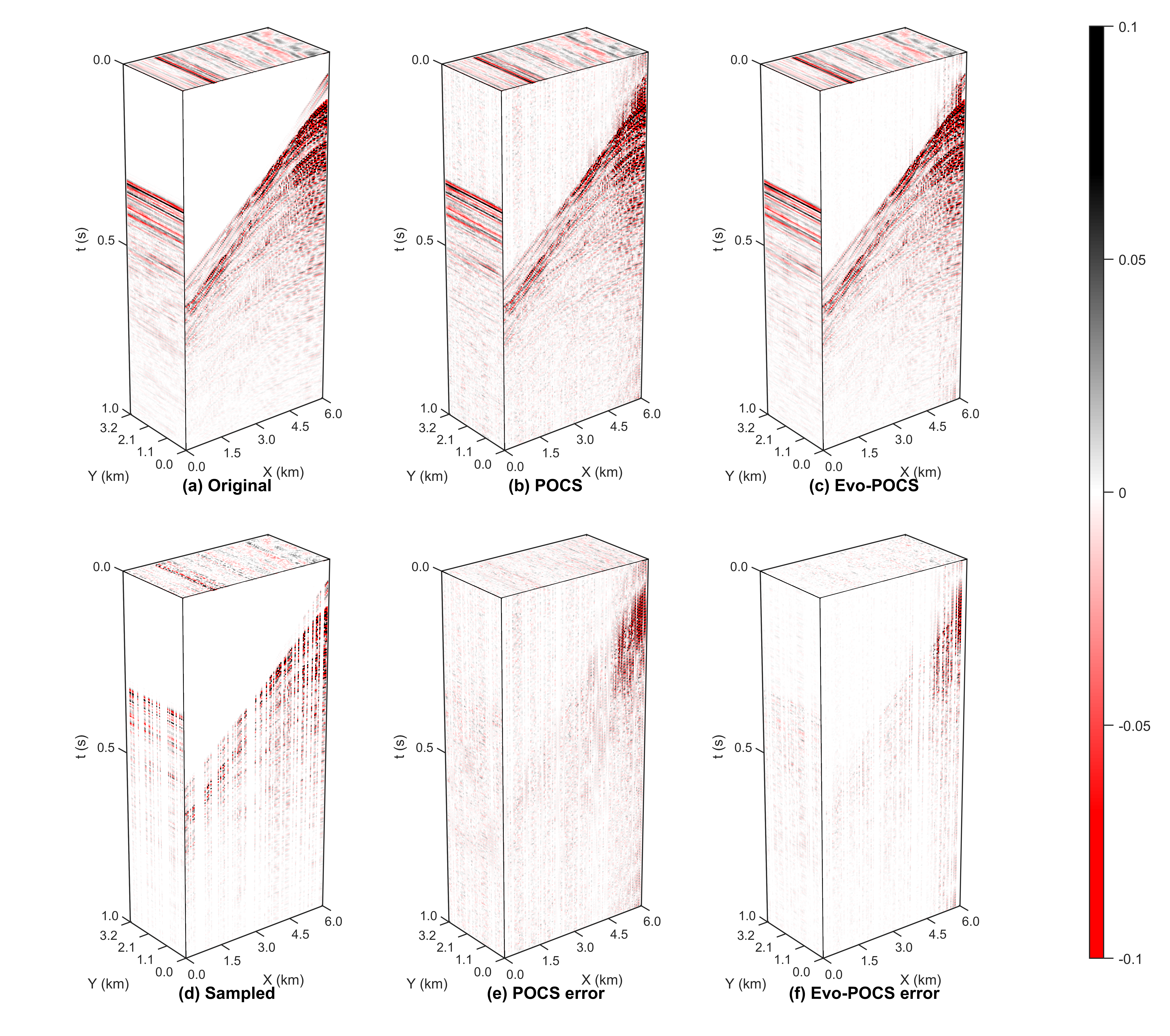}
   \caption{Reconstruction comparison on a $256\times120\times64$ field dataset
	under 50\% random sampling. (a) Original; (b) POCS; (c) Evo-POCS;
	(d) sampled data; (e) POCS error; (f) Evo-POCS error. The Evo-POCS
	reconstruction error is smaller than that of classic POCS.}
	\label{fig:avo12_rec_compare}
\end{figure}

\textbf{Generalization test.} To further assess the generalization of Evo-POCS,
we test reconstruction on several out-of-distribution synthetic and field
datasets. On these data, which were not used during the search, Evo-POCS
improves over classic POCS by more than 3~dB on average.
Figure~\ref{fig:avo12_rec_compare} shows the reconstruction on one field
dataset ($256\times120\times64$) under 50\% random sampling. Evo-POCS again
gives clearly better event continuity and a smaller reconstruction error. The
improvement is especially visible above the first arrivals, where Evo-POCS
suppresses the noise introduced by classic POCS and markedly reduces the error
in that region. Quantitatively, the SNR improves from 9.23~dB for classic POCS
to 12.79~dB, and SSIM from 0.980 to 0.992. 

\begin{figure}[htbp]
	\centering
	\includegraphics[width=1\textwidth]{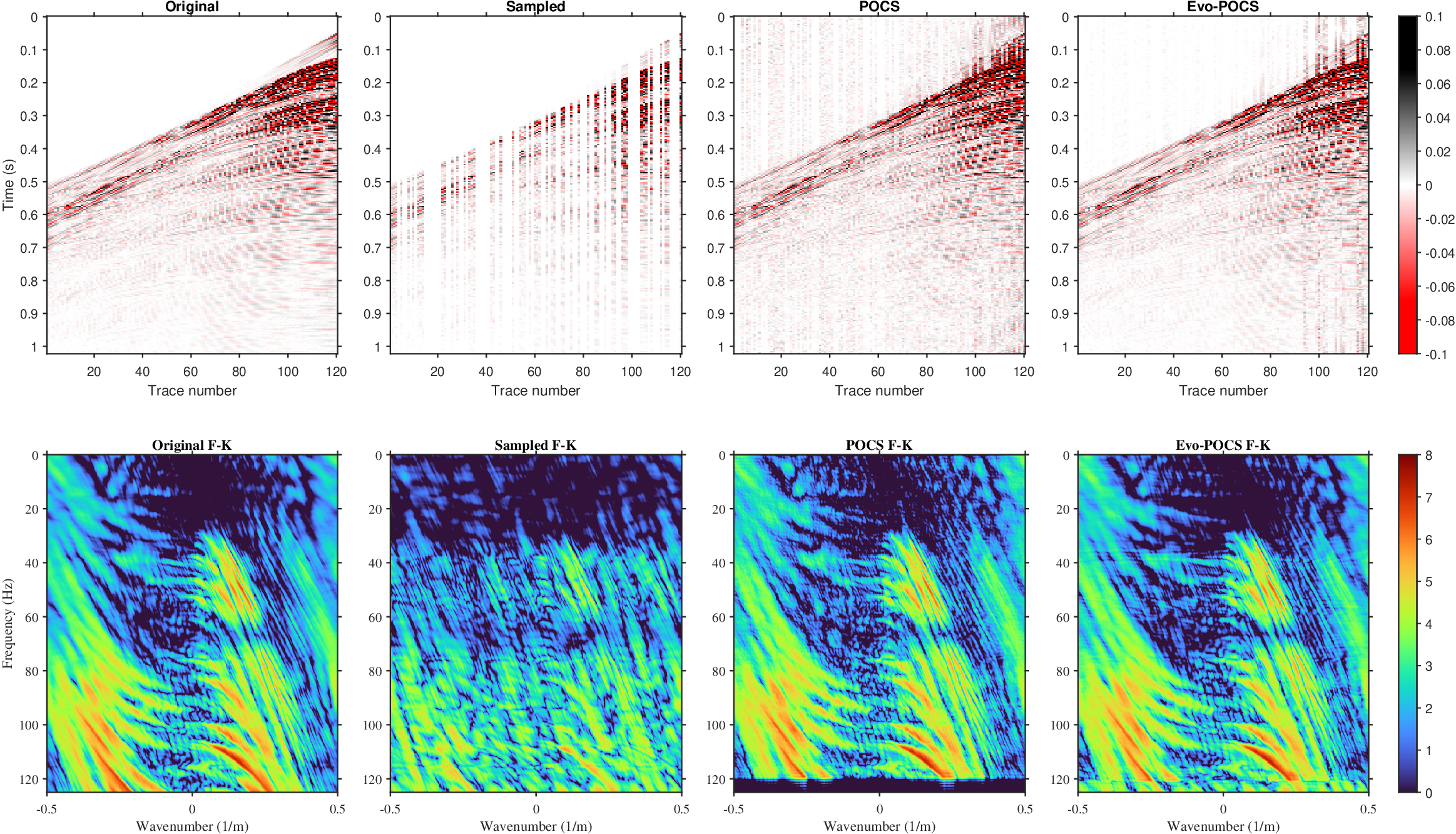}
	\caption{Reconstruction and $F$-$K$ comparison on a slice at $Y=0.65$~km.
			(a)--(d) original, sampled, POCS, and Evo-POCS reconstructions; (e)--(h) their
			$F$-$K$ spectra. Evo-POCS gives better event continuity and an $F$-$K$ spectrum
			closer to the original.}
	\label{fig:avo12_slice_compare}
\end{figure}

Figure~\ref{fig:avo12_slice_compare} further shows the reconstruction on a slice at $Y=0.65$~km together with the corresponding $F$-$K$ spectra. Both methods reconstruct the data effectively, but the advantage is clearer in this slice: Evo-POCS gives stronger event continuity and less noise, and
its $F$-$K$ spectrum is closer to that of the original data, with the sampling
aliasing better suppressed.

We further evaluate the sensitivity to the missing ratio on the same field
dataset. Table~\ref{tab:avo12_missing_snr} reports the SNR comparison between
POCS and Evo-POCS for missing ratios from 30\% to 70\%. Evo-POCS
consistently outperforms POCS at all missing ratios, with SNR gains from
1.58 to 5.42~dB and an average gain of 3.49~dB. This result shows that the
improvement is not limited to a single sampling density. The SNR gain gradually
decreases as the missing ratio becomes more severe. This behavior is expected
because the evolved dip-consistency correction relies on the structural evidence carried by the observed traces; when fewer traces are available, the
local dip information becomes less reliable and the prior has weaker support.
Nevertheless, Evo-POCS still improves over POCS even when 70\% of the traces are missing.
\begin{table}[t]
	\centering
	\caption{SNR comparison between classic POCS and Evo-POCS on the field
		dataset under different missing-trace ratios. $\Delta$SNR is computed as
		$\mathrm{SNR}_{\mathrm{Evo\text{-}POCS}} -
		\mathrm{SNR}_{\mathrm{POCS}}$.}
	\label{tab:avo12_missing_snr}
	\begin{tabular}{cccc}
		\toprule
		Missing ratio (\%) & POCS (dB) & Evo-POCS (dB) & $\Delta$SNR (dB) \\
		\midrule
		30 & 9.62 & \textbf{15.04} & + 5.42 \\
		40 & 9.48 & \textbf{13.73} & + 4.25 \\
		50 & 9.23 & \textbf{12.79} & + 3.56 \\
		60 & 8.88 & \textbf{11.52} & + 2.64 \\
		70 & 8.41 & \textbf{9.99}  & + 1.58 \\
		\midrule
		Average & 9.12 & \textbf{12.61} & + 3.49 \\
		\bottomrule
	\end{tabular}
\end{table}

%% file: sections/03_experiments2.tex
\subsection{SeisEvo test on MSSA for simultaneous interpolation and denoising}

\textbf{Task instantiation.} The second case study instantiates SeisEvo on the
MSSA family for simultaneous interpolation and denoising of 3D seismic data.
The observations follow the noisy sampling model in
Eq.~\eqref{eq:observation}, $Y=M\odot(X+N)$, where $M$ is a random spatial
sampling mask that retains 50\% of the complete traces. The entries of $N$ are
independent zero-mean Gaussian variables, $N\sim\mathcal N(0,\sigma^2)$, with
$\sigma=\alpha\max|X|$ and $\alpha\sim\mathcal U(0.01,0.20)$. Thus, the noise
level is scaled to the amplitude of each data block and varies across the
search cases. As in the POCS case study, the search uses 100 seismic sub-blocks
of size $64\times64\times64$. Candidate algorithms are generated and scored
only on these sub-blocks, and the external test data are not used during the
search.

The seed $A_0$ is a fixed-window implementation of the classical block-Hankel
MSSA low-rank model. The editable surface $\mathcal S$ includes singular-value
processing and rank selection, the Hankel low-rank projection, and the rule
used to estimate observation reliability in the soft data update. The temporal
Fourier transform, Hankel embedding, overlap averaging, input--output
interface, and mask-based form of the data update remain fixed. The constraint
set $\mathcal C$ requires the sampling mask to remain unchanged and the output
to be finite and real-valued. It also forbids test-set leakage and unexplained
free parameters.

Unlike the POCS case without added synthetic noise, this joint task assigns different roles to
the two data subsets: the acquired traces require denoising, whereas the
missing traces require interpolation. We therefore separate the SNR-related
assessment into the observed-trace denoising gain $G_{\mathrm{obs}}$ and the
missing-trace reconstruction SNR $\mathrm{SNR}_{\mathrm{miss}}$, while SSIM
measures structural fidelity over the complete reconstruction:
\begin{equation}
	O_{\mathrm{MSSA}}
	=
	0.35\,\operatorname{clip}
	\left(
	\frac{G_{\mathrm{obs}}}{18},0,1
	\right)
	+
	0.35\,\operatorname{clip}
	\left(
	\frac{\mathrm{SNR}_{\mathrm{miss}}}{18},0,1
	\right)
	+
	0.30\,\mathrm{SSIM},
	\label{eq:mssa_objective}
\end{equation}
where $G_{\mathrm{obs}}$ is the SNR gain of the reconstructed observed traces
over the noisy input, and $\mathrm{SNR}_{\mathrm{miss}}$ is evaluated only on
the missing traces. The 18~dB reference scales the two SNR terms to ranges
comparable with SSIM during the search; it is not a prescribed reconstruction
target.

\textbf{MSSA seed.} In practical seismic processing, MSSA is often
applied within local spatial windows. Local processing limits the size of the
block-Hankel matrix and reduces the cost of repeated low-rank decomposition. It
also makes the low-rank approximation more appropriate because seismic events
are closer to locally linear within a limited spatial aperture
\citep{wu2018adaptive}. Following this practice, we use a computationally
compact, fixed-window implementation of classical block-Hankel MSSA
\citep{oropeza2011simultaneous} as the seed. Here, ``compact'' refers only to
the use of a small local Hankel embedding. Throughout this paper, we denote
this seed simply as MSSA.

After a temporal Fourier transform, each frequency slice is processed
independently. For the $k$th frequency slice $U_k^{(j)}$, its low-rank estimate
is
\begin{equation}
	L_k^{(j)}
	=
	\mathcal H^\dagger
	\left[
	\mathcal P
	\left(
	\mathcal H\!\left(U_k^{(j)}\right)
	\right)
	\right],
	\label{eq:mssa_low_rank_step}
\end{equation}
where $\mathcal H$ is the spatial block-Hankel embedding, $\mathcal P$ is the
low-rank projection, and $\mathcal H^\dagger$ maps the projected Hankel matrix
back to the spatial frequency slice by overlap averaging. The MSSA seed and
all candidate algorithms use a fixed $6\times6$ Hankel window, a rank limit of
16, and 30 reconstruction iterations.

Because the observed traces are noisy, the seed does not restore them exactly.
Instead, it applies the soft data-consistency update
\begin{equation}
	U_k^{(j+1)}
	=
	M\odot
	\left[
	\gamma_{k,j}Y_k+
	\left(1-\gamma_{k,j}\right)L_k^{(j)}
	\right]
	+
	(1-M)\odot L_k^{(j)},
	\label{eq:mssa_soft_data_step}
\end{equation}
where $\gamma_{k,j}\in[0,1]$ balances the noisy observation $Y_k$ and the
low-rank estimate $L_k^{(j)}$. This implementation retains the central low-rank
model of classical MSSA while allowing the acquired traces to be corrected
during denoising.

\begin{figure}[H]
	\centering
	\includegraphics[width=0.6\textwidth]{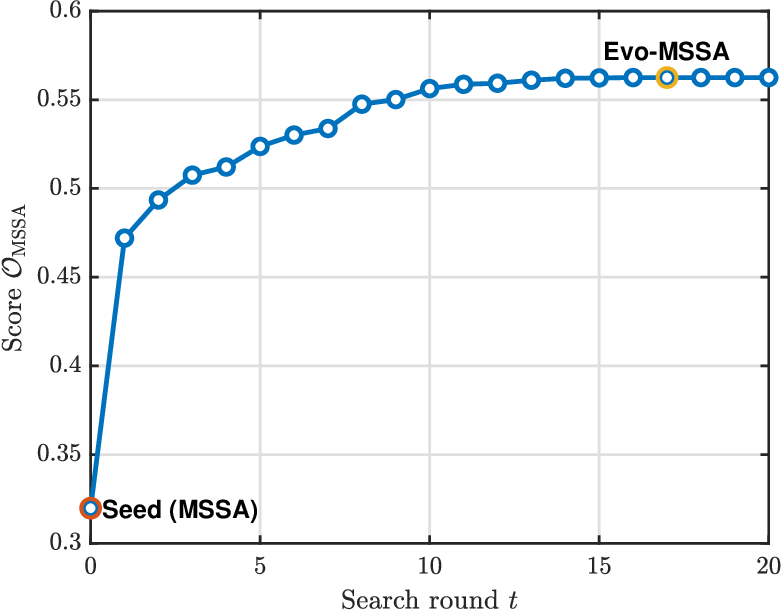}
	\caption{Search-time objective $O_{\mathrm{MSSA}}$ over 20 search rounds.
		The score increases from 0.320 for MSSA to 0.563 for
		the best candidate.}
	\label{fig:mssa_score}
\end{figure}

\textbf{Search process.} Starting from the MSSA seed, SeisEvo runs for
20 search rounds. Figure~\ref{fig:mssa_score} shows the search-time objective
$O_{\mathrm{MSSA}}$ over the search. The score increases from 0.320 for the
seed to 0.563 for the best candidate. Correspondingly, the observed-trace
denoising gain $G_{\mathrm{obs}}$ increases from 3.06 to 7.71~dB, the
missing-trace reconstruction SNR $\mathrm{SNR}_{\mathrm{miss}}$ from 3.83 to
8.59~dB, and the overall SSIM from 0.619 to 0.819. Although the overall
reconstruction SNR is not included directly in the search objective, it also
increases from 4.37 to 9.17~dB. These concurrent gains show that the search
improves observed-trace denoising, missing-trace interpolation, and structural
fidelity together, rather than improving one component at the expense of
another.

\textbf{Evo-MSSA.} The main structural change found by SeisEvo is a
reliability-grouped singular-value shrinkage. In one sentence, Evo-MSSA is MSSA
with data-driven patch grouping: rather than fitting all Hankel patches with
one fixed-rank subspace, it groups them by local observation reliability and
fits each group with its own low-rank subspace. The fixed Hankel window, rank
limit, and reconstruction count remain unchanged from the MSSA seed.

Algorithm~\ref{alg:evomssa} summarizes the deployed operator for one frequency
slice. The noisy observation $Y_k$ is fixed throughout the iterations and
serves as both the initial value and the anchor of the soft data update. In the
first iteration, when no previous reliability field is available, Evo-MSSA
applies a global MP--GD shrinkage. Here, MP--GD denotes a singular-value
shrinkage whose noise scale is estimated from the lower spectral tail and
calibrated by the Marchenko--Pastur law, while its threshold coefficient is
given by the Gavish--Donoho optimal hard-threshold form
\citep{marchenko1967distribution,gavish2014optimal}.

From the second iteration onward, the Hankel columns are divided at the
empirical mean of the patch reliability derived from
$\Gamma_k^{(j-1)}$. Each group is projected using its own singular spectrum
and matrix aspect ratio, and the projected columns are placed back in their
original positions. The resulting low-rank slice is then compared with $Y_k$
on the observed traces. Overlapping-support averaging and a centered
cross-iteration Ces\`aro average convert these residuals into the current
reliability field $\Gamma_k^{(j)}$, which controls the soft update. The complete
MP--GD, grouping, reliability, and fallback expressions are given in
Appendix~\ref{sec:appendix_evomssa}.

\begin{algorithm}[H]
	\caption{Evo-MSSA reconstruction for one frequency slice}
	\label{alg:evomssa}
	\begin{algorithmic}[1]
		\Require fixed noisy observed slice $Y_k$; mask $M$;
		iterations $J$; Hankel operator $\mathcal H$
		\Ensure reconstructed slice $U_k$
		\State $U_k^{(0)}\gets Y_k$;
		$\Gamma_k^{(-1)}\gets\varnothing$
		\For{$j=0,\ldots,J-1$}
		\State $\pi_j\gets j/(J-1)$
		\State $B_k^{(j)}\gets\mathcal H(U_k^{(j)})$
		\If{$j=0$}
		\State $\widetilde B_k^{(j)}
		\gets\operatorname{MPGD}(B_k^{(j)})$
		\Else
		\State $\widetilde B_k^{(j)}
		\gets\operatorname{GroupMPGD}
		(B_k^{(j)},\Gamma_k^{(j-1)})$
		\EndIf
		\State $L_k^{(j)}
		\gets\mathcal H^\dagger(\widetilde B_k^{(j)})$
		\State $\Gamma_k^{(j)}
		\gets\operatorname{ConsensusTrust}
		(Y_k,L_k^{(j)},M,\pi_j)$
		\State $U_k^{(j+1)}\gets
		M\odot[\Gamma_k^{(j)}\odot Y_k+
		(1-\Gamma_k^{(j)})\odot L_k^{(j)}]
		+(1-M)\odot L_k^{(j)}$
		\EndFor
		\State \Return $U_k^{(J)}$
	\end{algorithmic}
\end{algorithm}

The three modules in Algorithm~\ref{alg:evomssa} have distinct roles.
$\operatorname{MPGD}$ adapts the shrinkage to the measured spectrum and matrix
shape; $\operatorname{GroupMPGD}$ permits patches with different reliability
to use different subspaces; and $\operatorname{ConsensusTrust}$ stabilizes the
residual-based reliability estimate in space and across iterations. The
evolved mechanism introduces no manually tuned continuous parameter.
The two-group structure, empirical-mean split, robust residual scale, and MP
plug-in remain explicit structural choices rather than assumptions hidden in
learned weights.

\begin{figure}[H]
	\centering
	\includegraphics[width=1\textwidth]{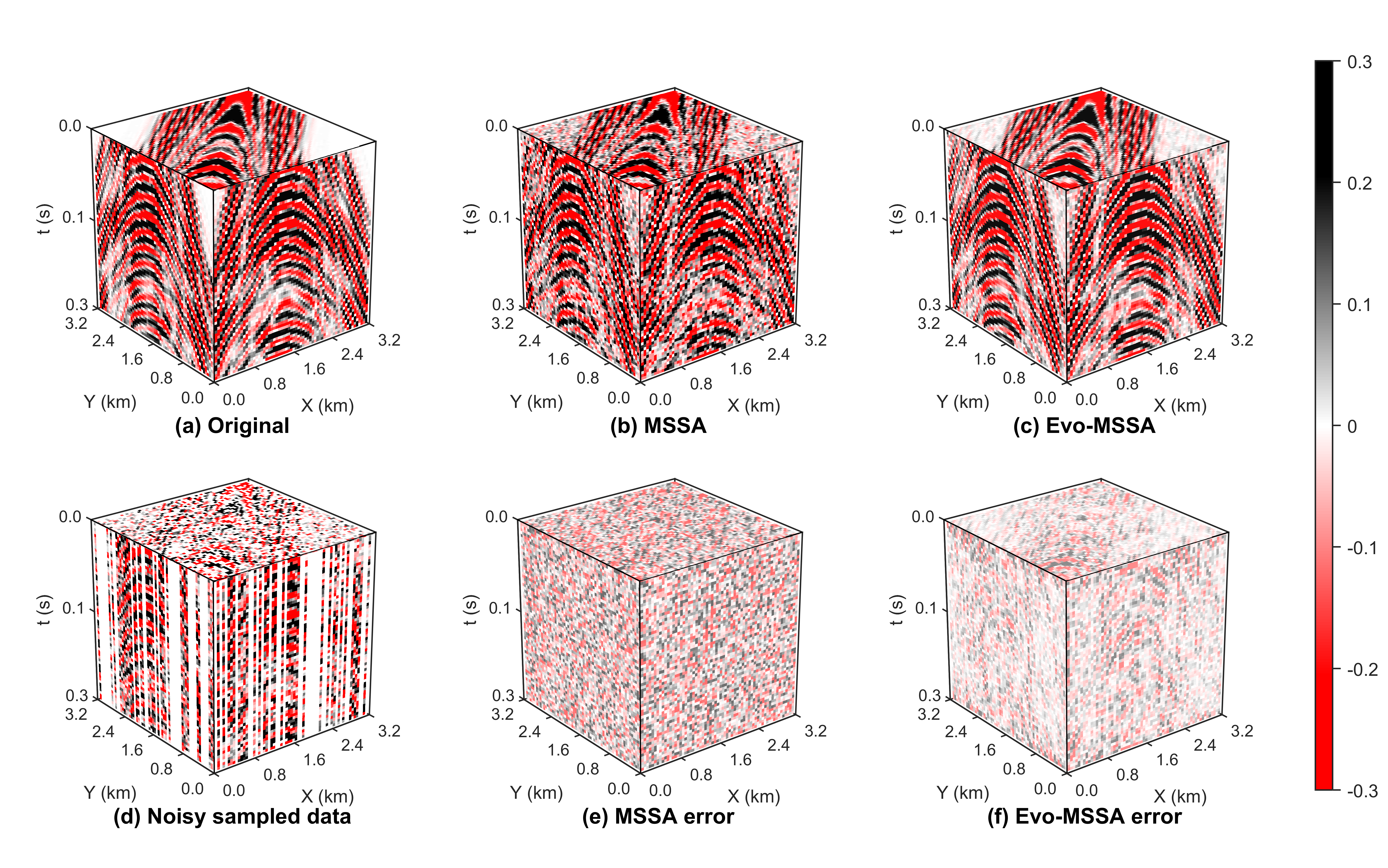}
	\caption{Reconstruction comparison on a $64\times64\times64$ synthetic
		sub-block with 50\% randomly missing traces and Gaussian noise.
		(a) Original complete data; (b) MSSA reconstruction;
		(c) Evo-MSSA reconstruction; (d) noisy data with 50\% randomly missing
		traces; (e) MSSA error; (f) Evo-MSSA error.}
	\label{fig:syn_evomssa_compare}
\end{figure}

\textbf{Reconstruction results.} Figure~\ref{fig:syn_evomssa_compare} compares
MSSA and Evo-MSSA on a $64\times64\times64$ synthetic sub-block with 50\%
randomly missing traces and Gaussian noise. Panel (a) shows the original
complete data, panel (d) the noisy incomplete data, panels (b) and (c) the
reconstructions obtained by MSSA and Evo-MSSA, and panels (e) and (f) their
corresponding errors. Both methods recover the main seismic events across the
missing traces. However, visible residual noise remains in the MSSA result.
Evo-MSSA produces more continuous events, weaker background noise, and a
smoother reconstruction. The error panels further show that Evo-MSSA leaves
less residual noise and a smaller overall reconstruction error.

The input SNR of the noisy incomplete data is 1.91~dB. MSSA increases the SNR
to 7.03~dB, whereas Evo-MSSA reaches 11.09~dB, an improvement of 4.06~dB over
MSSA. The MSE is also reduced from $6.74\times10^{-3}$ for MSSA to
$2.65\times10^{-3}$ for Evo-MSSA. These results show that the Evo-MSSA
operator obtained through the SeisEvo search clearly improves on the MSSA
seed. It recovers events across missing traces while more effectively
attenuating noise on the observed traces.

\textbf{Generalization test.} To further assess the generalization of
Evo-MSSA, we test simultaneous interpolation and denoising on several
synthetic and field datasets that were not used during the search. Across
these unseen datasets, Evo-MSSA improves the overall reconstruction SNR over
MSSA by more than 4~dB on average. This result shows that the mechanism found
by SeisEvo is not limited to the sub-blocks used during the search and can
generalize to unseen synthetic and field seismic data.

\begin{figure}[H]
	\centering
	\includegraphics[width=1\textwidth]{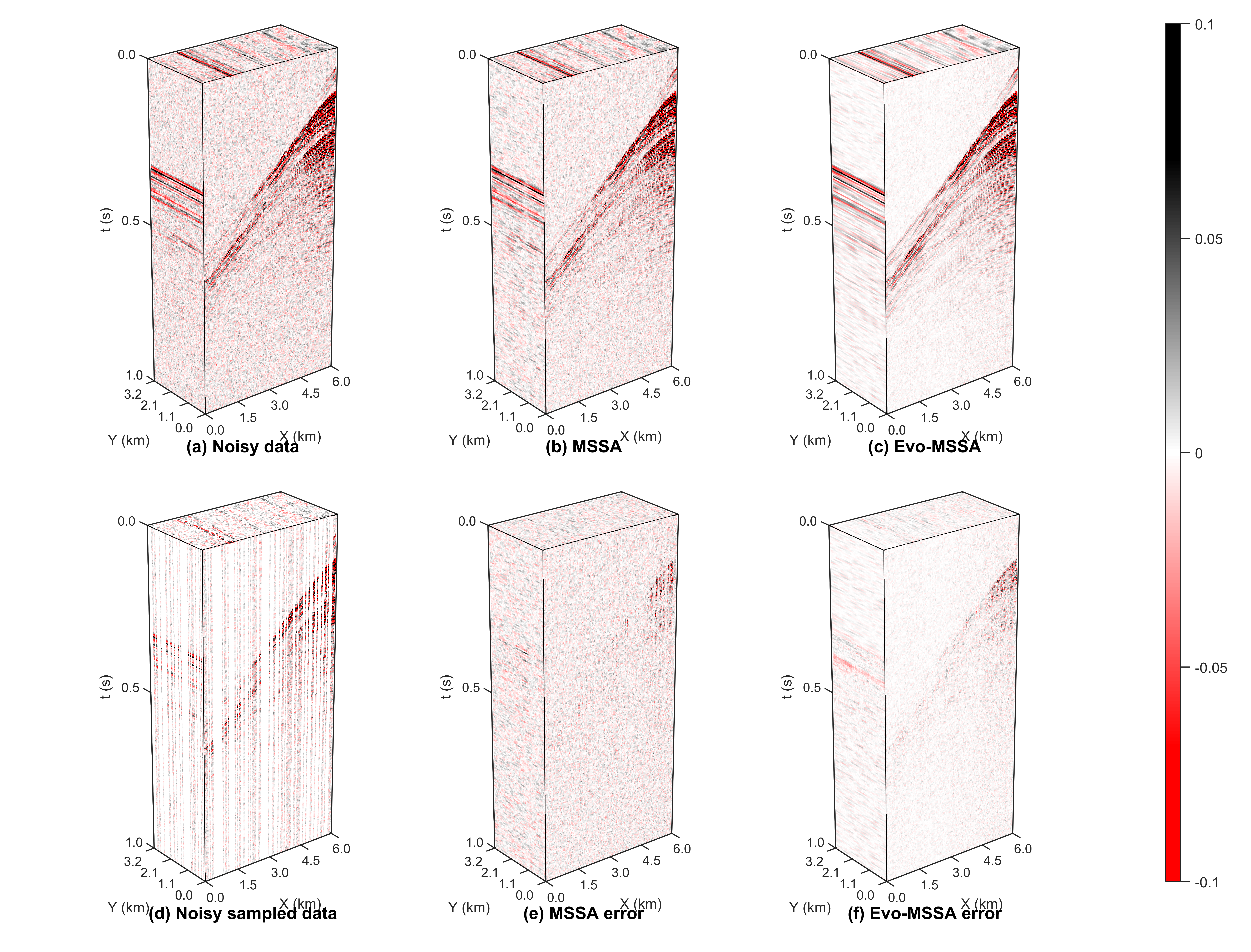}
	\caption{Reconstruction comparison on a $256\times120\times64$ field dataset
		with 50\% randomly missing traces and Gaussian noise. (a) Noisy complete data;
		(b) MSSA reconstruction; (c) Evo-MSSA reconstruction; (d) noisy incomplete data; (e) MSSA error;
		(f) Evo-MSSA error.}
	\label{fig:avo12_mssa_rec_compare}
\end{figure}

Figure~\ref{fig:avo12_mssa_rec_compare} shows a representative result on a
$256\times120\times64$ field dataset outside the search set. Panel (a) shows
the complete data after noise injection, whereas panel (d) shows the same noisy
data after 50\% of the traces are removed. The SNR of the
noisy incomplete input is 2.95~dB. MSSA increases the reconstruction SNR to
10.88~dB, whereas Evo-MSSA reaches 14.49~dB, a gain of 3.61~dB over MSSA.
The corresponding MSE decreases from $9.84\times10^{-5}$ for MSSA to
$3.92\times10^{-5}$ for Evo-MSSA. Both methods recover the main seismic
events, but the Evo-MSSA result is smoother and shows stronger noise
attenuation. The difference is clearer in the error panels: visible residual
noise remains after MSSA reconstruction, whereas Evo-MSSA produces a weaker
and cleaner reconstruction error.

\textbf{Comparison with ODRR.} To determine
whether the improvement extends beyond the seed method, we further compare
Evo-MSSA with the optimally damped rank-reduction method (ODRR)
\citep{chen2020odrr,chen2023drr}. Following the recommended parameter range,
a small validation sweep selected $K=3$. Except for the damping setting
specific to ODRR, the reconstruction
parameters shared by the methods were kept unchanged.

We vary the noise scale $\alpha$ from 0.01 to 0.05 on the same field dataset with 50\% randomly missing traces.
Table~\ref{tab:odrr_noise_snr} reports the overall reconstruction SNR.
Evo-MSSA achieves the highest SNR at every noise level, and its advantage
becomes more pronounced at higher noise levels. Averaged over the five noise
levels, Evo-MSSA reaches 10.70~dB, exceeding MSSA and ODRR by 7.15 and
3.31~dB, respectively. These results show that SeisEvo improves the classical
MSSA seed and remains effective relative to the ODRR baseline.

\begin{table}[t]
	\centering
	\caption{Overall reconstruction SNR on the field dataset under different
		noise scales.}
	\label{tab:odrr_noise_snr}
	\begin{tabular}{ccccc}
		\toprule
		Noise scale $\alpha$ & MSSA (dB) & ODRR (dB) & Evo-MSSA (dB) &
		\makecell{Gain over\\ODRR (dB)} \\
		\midrule
		0.01 & 10.88 & 12.61 & \textbf{14.49} & $+1.88$ \\
		0.02 & 5.77  & 8.84  & \textbf{12.05} & $+3.21$ \\
		0.03 & 2.58  & 7.13  & \textbf{10.32} & $+3.19$ \\
		0.04 & 0.23  & 5.50  & \textbf{8.96}  & $+3.46$ \\
		0.05 & -1.71 & 2.85  & \textbf{7.70}  & $+4.85$ \\
		\midrule
		Average & 3.55 & 7.39 & \textbf{10.70} & $+3.31$ \\
		\bottomrule
	\end{tabular}
\end{table}

\begin{figure}[H]
	\centering
	\includegraphics[width=1\textwidth]{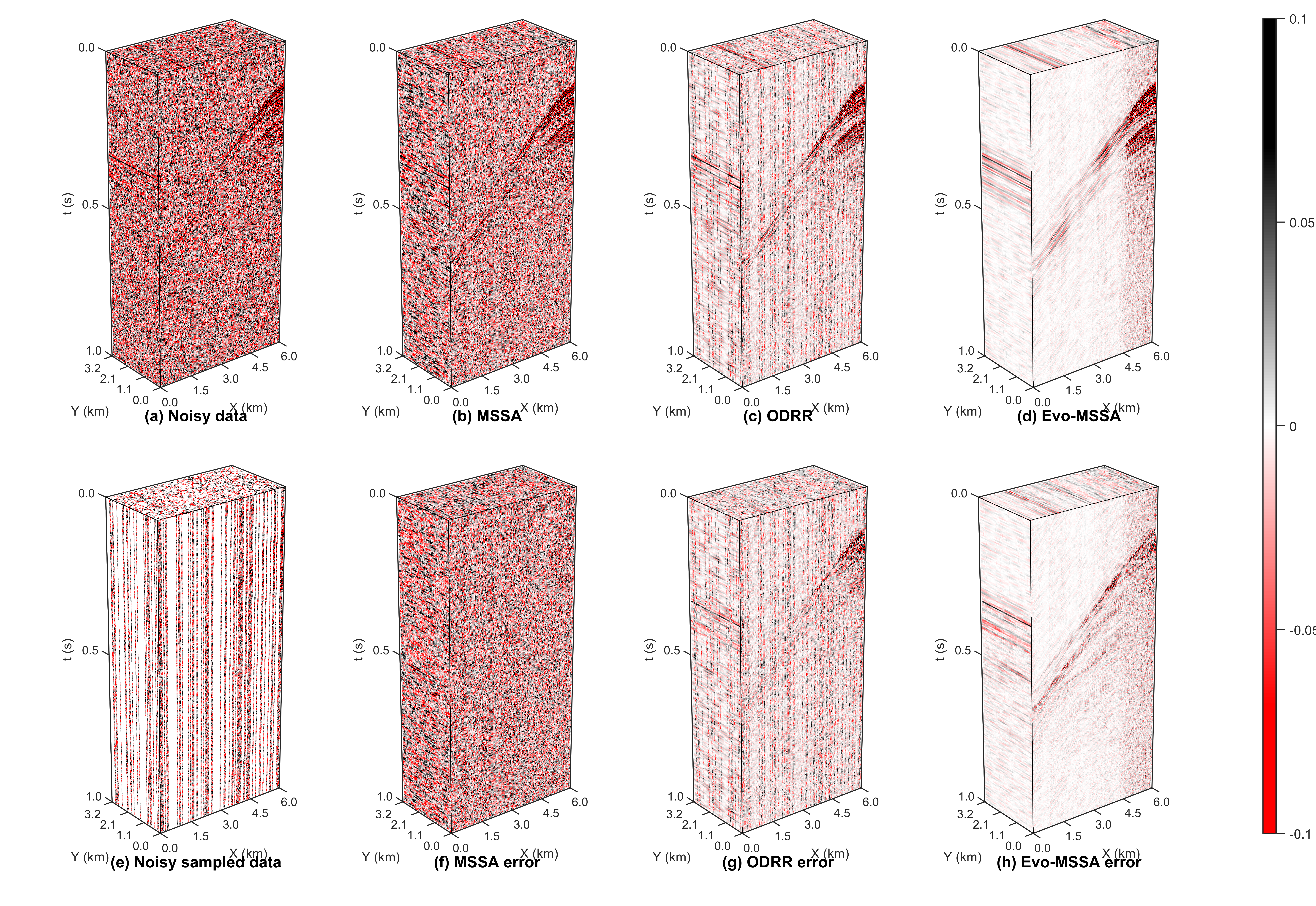}
	\caption{Reconstruction comparison on a $256\times120\times64$ field dataset
		at noise scale $\alpha=0.05$ with 50\% randomly missing traces.
		(a) Noisy complete data; (b) MSSA reconstruction;
		(c) ODRR reconstruction; (d) Evo-MSSA reconstruction;
		(e) noisy sampled data; (f) MSSA error; (g) ODRR error;
		(h) Evo-MSSA error.}
	\label{fig:avo12_mssa_comare_diffnoise}
\end{figure}

Figure~\ref{fig:avo12_mssa_comare_diffnoise} provides a visual comparison at
$\alpha=0.05$. Under this low-input-SNR condition, simultaneous interpolation
and denoising are difficult. The MSSA reconstruction remains
dominated by noise, and most seismic events are barely identifiable. ODRR
recovers part of the event structure, but substantial residual noise and
spatial artifacts remain. Evo-MSSA gives the clearest result: the main events
remain visible and show better continuity. Nevertheless, its error volume
still contains coherent signal energy and random noise, indicating signal
leakage and residual noise. This comparison shows that Evo-MSSA is more robust
than MSSA and ODRR under this low-input-SNR condition, while also revealing its
limitations in this challenging case.

\begin{figure}[H]
	\centering
	\includegraphics[width=1\textwidth]{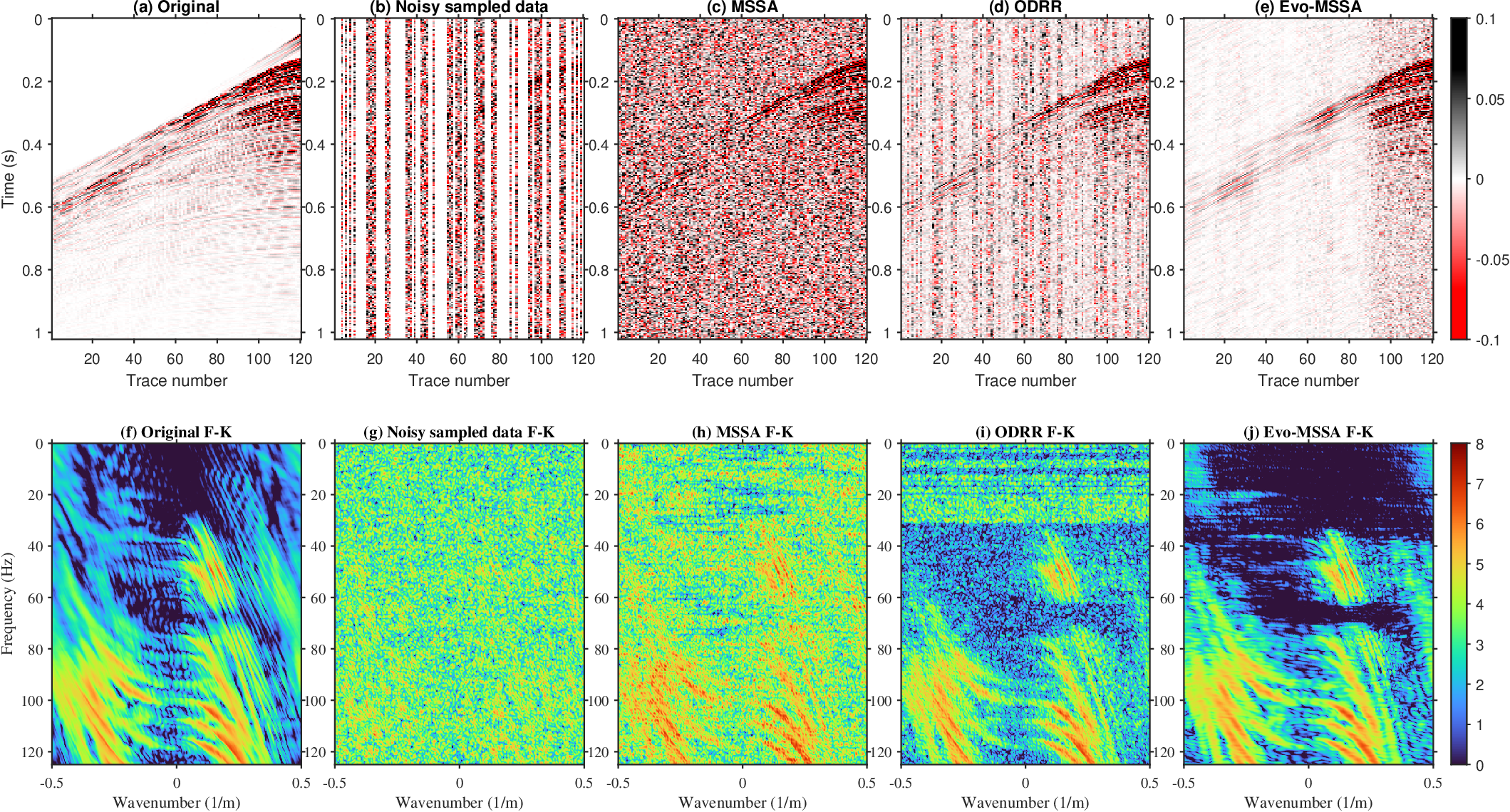}
	\caption{Slice and $F$--$K$ spectrum comparison on a
		$256\times120\times64$ field dataset at $Y=1.75$~km, with noise
		scale $\alpha=0.05$ and 50\% randomly missing traces.
		(a) Original; (b) noisy sampled data; (c) MSSA;
		(d) ODRR; (e) Evo-MSSA; (f)--(j) the corresponding
		$F$--$K$ spectra of (a)--(e), respectively.}
	\label{fig:avo12_mssa_compare_slice_fk2}
\end{figure}

Figure~\ref{fig:avo12_mssa_compare_slice_fk2} further compares the slice at
$Y=1.75$~km and the corresponding $F$--$K$ spectra. At $\alpha=0.05$, random
noise and missing traces make the original events almost unidentifiable in
the noisy sampled data. Its $F$--$K$ spectrum also contains little
identifiable signal energy. MSSA does not effectively separate the signal
from noise: its reconstruction remains dominated by noise, and the
corresponding $F$--$K$ spectrum shows the same behavior. ODRR recovers part of
the main events and their spectral structures, but substantial residual noise
remains.

In comparison, Evo-MSSA recovers clearer and more continuous main events and
further suppresses the effects of random noise and missing traces. Its
$F$--$K$ energy is more concentrated around the main signal structures and is
closer to that of the original data. However, some weak events and spectral
energy are still attenuated, indicating remaining signal loss under this
low-input-SNR condition. The slice comparison is consistent with the
full-volume results and quantitative metrics, further showing the improved
interpolation and denoising performance of Evo-MSSA at low SNR.

Taken together, the two case studies start from Fourier-sparse POCS and
low-rank MSSA and yield Evo-POCS and Evo-MSSA, respectively. In the POCS case,
Evo-POCS retains the Fourier sparse projection and data-consistency constraint
while mainly adding a data-driven dip-consistency correction, which improves
the continuity of reconstructed events. In the MSSA case, Evo-MSSA improves
the conventional low-rank reconstruction through adaptive singular-value
shrinkage, reliability grouping, and cross-iteration consensus. It also
maintains an advantage over ODRR, especially at higher
noise levels.

The two cases address different reconstruction tasks: Evo-POCS considers
interpolation of randomly missing traces without added synthetic noise, whereas Evo-MSSA
considers simultaneous interpolation and denoising with randomly missing traces
and Gaussian noise. In their respective tasks, both evolved algorithms clearly
outperform their classical seeds and maintain consistent improvements on data
outside the search set. These results indicate that the mechanisms discovered
by SeisEvo can work with algorithm families based on different mathematical
principles and under different observation conditions and reconstruction
objectives, rather than being limited to the sub-blocks used during search.
Overall, the experiments support the effectiveness of SeisEvo on two
classical seismic reconstruction families and two different reconstruction
tasks.

%% file: sections/04_discussion_conclusion.tex
\section{Discussion}

\paragraph{What the search actually contributed.}
The discovered mechanisms are not conceptually new. Interpolation that exploits
dip continuity, and adaptive shrinkage of singular values, each have their own
research background in seismic processing. The value of SeisEvo does not lie in
proposing a previously unseen prior, but in the fact that the specific form of
these mechanisms was constructed by the search without the form of the operator
being prescribed, and was integrated automatically into the iteration of a
classical solver. In Evo-POCS, the directional candidate set, the phase
estimation from observed trace pairs, the residual-weighted directional
consensus, and the confidence gating were not given by the specification; in
Evo-MSSA, neither was the structure that divides the Hankel columns by local
reliability and projects each group with its own subspace. These elements were
built up and selected under the physical-legality and parameter-provenance
constraints. The resulting operators remain explicit algorithms that can be
written down, inspected at the operator level, related to interpretable seismic
mechanisms, and tested independently. This is what distinguishes algorithm
evolution from replacing a solver with an implicit mapping: it changes the way
an algorithm is designed, not the way an algorithm is expressed.

\paragraph{Scope and cost of the protocol.}
More generally, the conditions on which SeisEvo relies, namely an executable
quality measure, explicit physical contracts, and a classical algorithm with a
clear and modular structure, are also present in seismic denoising,
deconvolution, and inversion, so the protocol may extend to other steps of
seismic signal processing. The main cost at present is offline search
computation: discovering an operator requires many candidate executions and is
therefore substantially more expensive than running a fixed hand-designed
solver. This cost is paid only during discovery; once frozen, the operator requires no
agent, prompt, or neural network at inference time. The present
evidence, however, is restricted to seismic data reconstruction; extension to
these tasks remains a hypothesis to be tested rather than a demonstrated
capability.

\paragraph{Design of the objective and evaluator.}
This work makes the constraint set and the editable operator surface explicit
and part of the search protocol, but the seed and the objective are fixed
before each run, and the effect of their choice has not been studied
systematically. The objective encodes part of the human judgement of quality as
an executable scoring rule. The two case studies use different scoring
structures: the POCS search combines the reconstruction SNR with the structural
similarity, whereas the MSSA search further separates noise suppression at the
observed positions from reconstruction quality at the missing ones. The
associated weights and normalizing references belong to an evaluation protocol
that is identical for all candidates; they are fixed before the search and held
by the evaluator, and no candidate program may modify them. The objective plays
a role comparable to the loss function of a learning method, in that both
prescribe the optimization preference, but it guides a search over discrete
programs and operator mechanisms rather than the optimization of network
weights. It therefore need not be differentiable and may include piecewise
measures and discrete quality judgements, while legality checks are enforced
separately through the constraint set. Different objectives consequently create
different search pressures: raising the SNR, preserving structure, suppressing
noise, and limiting signal leakage are not always mutually consistent. An
executable evaluator thus does more than record performance; it defines the
selection pressure under which a program evolves
\citep{novikov2025alphaevolve}. The executable form of the objective can itself
become an object of search, as in the automatic design and evolution of reward
code \citep{ma2024eureka}. In the present runs, the best-so-far scores exhibit
late-stage plateaus, but a single trajectory per case is not sufficient to
decide whether the budget is already adequate, whether candidate diversity is
being exhausted, or whether the room for observable improvement under the
current evaluation protocol has simply become small. Separating these
possibilities would require repeated independent searches, controlled
experiments with longer budgets, and explicit stopping rules based on
performance gain, candidate diversity, and computational cost.

\paragraph{From offline discovery to continual discovery.}
The SNR and the structural similarity used here both require a complete
reference, so the present protocol suits offline algorithm discovery when a
complete reference is available or when degraded observations can be
constructed from complete data, whereas the traces that are genuinely missing
in an acquired survey are themselves unobservable. For interpolation, part of
the observed traces can be hidden and used for self-supervised validation; for
simultaneous interpolation and denoising, the hidden traces are themselves
noisy, so additional assumptions about noise independence, repeated
observations, or noise statistics are needed. Spectral consistency, local
coherence, and the quality of subsequent imaging can also serve as substitutes,
but they remain proxy objectives and may introduce their own bias into the
search, which is consistent with the preceding observation that the scoring
design shapes the mechanism that is found. A separate issue is the dependence
on the seed and the search backend. Each case study here examines a single
classical seed with a fixed backend; whether different seeds would arrive at
comparable mechanisms, and how far objectives and constraints with more
specific geophysical meaning, such as band preservation, amplitude fidelity, or
a forward-modelling residual, would shape the direction of the search, remain
to be examined. If the discovery process is to continue, for instance by taking
a discovered operator as the seed of a further run, then repeatedly evaluating
against the same external data would gradually make that data part of the
search and cost it its role as independent validation. Beyond this, what
evolves in this work is the reconstruction solver rather than the search
mechanism itself. Adaptive optimization of candidate generation, experience
organization, and budget allocation is a natural extension: as long as the
evaluator and the hard constraints stay fixed, the legality decisions and the
scores of executed candidates remain reproducible under fixed programmatic
rules. If the objective or the legality rules were to co-evolve as well,
however, the standard of evaluation would become a moving target, and a fixed
outer validation protocol that does not participate in the evolution would be
required; otherwise an increase in score would no longer necessarily correspond
to an improvement of the algorithm itself.

\section{Conclusions}

We proposed SeisEvo (Seismic Algorithm Evolution), an algorithm evolution
framework for seismic data reconstruction. Rather than optimizing a single
reconstructed volume, SeisEvo searches for the reconstruction algorithm itself.
A domain expert defines a classical seed algorithm, the components that may be edited, the reconstruction objective, and the legality constraints; an LLM-based multi-agent search then improves executable candidates within those bounds. The final
output is a standalone white-box operator that requires no agent, prompt, or
neural network at inference time.

We instantiated SeisEvo on two classical algorithm families with different
mathematical foundations. For interpolation of field data without added
synthetic noise, Evo-POCS improves the SNR over classic POCS by 3.49 dB on
average across missing ratios from 30\% to 70\%, with the gain decreasing from
5.42 dB at 30\% missing to 1.58 dB at 70\% missing. For simultaneous interpolation and denoising,
Evo-MSSA improves the average reconstruction SNR by more than 7 dB over classic
MSSA and by more than 3 dB over the ODRR baseline, under 50\% randomly missing
traces and across the tested noise levels. Both evolved operators maintain their
performance gains on synthetic and field data not used during the search.

These results suggest a different role for domain expertise in geophysical algorithm design: the expert defines what constitutes a legal and useful algorithm, while the mechanism itself is discovered through search, and the resulting operators remain explicit, inspectable, and directly deployable. SeisEvo therefore indicates the potential of
agentic algorithm evolution as a complementary path to deep learning for
discovering interpretable geophysical processing algorithms.

%% file: sections/05_acknowledgments.tex
\section*{Acknowledgments}

This work was supported in part by Deep Earth Probe and Mineral Resources Exploration - National Science and Technology Major Project under Grant No. 2024ZD1002700, in part by the NSFC under Grant No. 42574172. Yingjie Xu acknowledges the support of the China Scholarship Council program (CSC No. 202506120138).



%% file: sections/appendix.tex
\appendix
\renewcommand{\theequation}{\Alph{section}.\arabic{equation}}
\setcounter{equation}{0}

\section{Mathematical formulation of the dip-consistency correction in Evo-POCS}
\label{sec:appendix_evopocs}

This appendix gives the mathematical form of the dip-consistency correction in
Evo-POCS. All operations are performed on a single frequency slice. Let $Y_k$ be
the observed frequency slice used in Eq.~\eqref{eq:pocs_data_step}, $M$ the
sampling mask, $\mathbf U$ the current iterate, and $u=(x,y)$ a spatial
location. Shifted locations $u\pm d$ that fall outside the slice are excluded
from the local sums below.

\paragraph{Direction bank} The correction is applied over a fixed set of
candidate dip directions, each a small integer spatial offset:
\begin{equation}
	\begin{split}
		\mathcal D = \{&(1,0),(0,1),(1,1),(1,-1),(2,1),(1,2),\\
		&(2,-1),(1,-2),(2,0),(0,2),(2,2),(2,-2)\}.
	\end{split}
\end{equation}
These twelve offsets, with steps at most two, form a discrete set of local dip
directions for seismic events.

\paragraph{Local phase coefficient} For each direction $d=(d_x,d_y)$, a complex
phase coefficient $c_d(u)$ is estimated from observed trace pairs, describing the
phase rotation between neighboring traces along that direction. Within a
$7\times7$ window $\mathcal N_7(u)$ centered at $u$,
\begin{equation}
	c_d(u) =
	\frac{\sum_{v\in\mathcal N_7(u)} \overline{Y_k(v+d)}\,Y_k(v)\,M(v+d)M(v)}
	{\left|\sum_{v\in\mathcal N_7(u)} \overline{Y_k(v+d)}\,Y_k(v)\,M(v+d)M(v)\right|+\epsilon},
\end{equation}
where $\epsilon$ is a small constant that prevents division by zero. The
normalization makes $c_d(u)$ have approximately unit modulus, so it acts mainly
as a phase rotation; the mask product $M(v+d)M(v)$ ensures that only trace pairs
observed at both ends are used. Since $c_d(u)$ depends only on the observed
slice and the mask, it is computed once before the iterations.

\paragraph{Directional prediction and residual} Given the current iterate
$\mathbf U$, each direction predicts the current location by symmetric
phase-aligned continuation from its two neighbors:
\begin{equation}
	P_d(\mathbf U)(u) = \tfrac{1}{2}\left[c_d(u)\,\mathbf U(u+d) + \overline{c_d(u)}\,\mathbf U(u-d)\right].
\end{equation}
Indeed, if the local wavefield satisfies
$\mathbf U(u+d)=e^{\mathrm i\phi_d}\mathbf U(u)$, then
$c_d(u)\approx e^{-\mathrm i\phi_d}$ under Eq.~(A.2). Consequently, both
$c_d(u)\mathbf U(u+d)$ and $\overline{c_d(u)}\mathbf U(u-d)$ predict
$\mathbf U(u)$; the conjugate in the second term provides the reverse-direction
phase alignment.
The reliability of this prediction is measured by its residual on observed
samples, within a $3\times3$ window $\mathcal N_3(u)$:
\begin{equation}
	R_d(u) =
	\frac{\sum_{v\in\mathcal N_3(u)} M(v)\,|Y_k(v)-P_d(\mathbf U)(v)|^2}
	{\sum_{v\in\mathcal N_3(u)} M(v)+\epsilon}.
\end{equation}
A smaller residual indicates that the direction better matches the local dip of
the events.

\paragraph{Consensus prediction} Let $\bar R(u)$ be the mean residual over the
twelve directions, and let $d_1,d_2,d_3$ be the three directions with the
smallest residuals. They are combined into a consensus prediction using cubic
residual-margin weights:
\begin{equation}
	w_i(u) = \left[\bar R(u)-R_{d_i}(u)\right]_+^3,
	\qquad
	P_\star(\mathbf U)(u) = \frac{\sum_{i=1}^{3} w_i(u)\,P_{d_i}(\mathbf U)(u)}{\sum_{i=1}^{3} w_i(u)+\epsilon},
\end{equation}
where $[\cdot]_+$ takes the non-negative part. Directions with smaller residuals
receive larger weights, so the most self-consistent direction dominates the
prediction.

\paragraph{Confidence gate and update} The strength of the correction is
controlled by a confidence $\kappa(u)$, which measures how much the best
direction improves over the mean residual:
\begin{equation}
	\kappa(u) = \operatorname{clip}\!\left(\frac{\bar R(u)-R_{\min}(u)}{\bar R(u)+\epsilon},\,0,\,1\right),
	\qquad
	R_{\min}(u) = \min_{d\in\mathcal D} R_d(u).
\end{equation}
When one direction clearly outperforms the rest, $\kappa$ is close to one and
the correction is strong; when the residuals are similar, $\kappa$ is close to
zero and the correction is weak. Writing $\mathbf U^{\mathrm{POCS}}$ for the
output of the classic POCS sparse projection (i.e., $V_k^{(j)}$ in
Eq.~\eqref{eq:pocs_sparse_step}), one full Evo-POCS update is
\begin{equation}
	\mathbf U^{(j+1)} = Y_k + (1-M)\odot
	\left[\mathbf U^{\mathrm{POCS}} + \kappa\odot\left(P_\star(\mathbf U^{\mathrm{POCS}})-\mathbf U^{\mathrm{POCS}}\right)\right].
\end{equation}
The outer term $Y_k + (1-M)\odot(\cdot)$ restores the observed traces exactly,
so only the missing traces are updated by the correction.

\paragraph{Fallback rules} When a $7\times7$ window contains no usable observed
trace pair, $c_d(u)$ falls back to a global phase estimate for the slice; when a
$3\times3$ window contains no observed sample, the residual $R_d(u)$ is replaced
by a self-consistency measure across directions. These fallbacks help keep the
operator stable where observations are sparse.

\paragraph{Real output} After all slices are processed, Hermitian symmetry is
imposed before the inverse temporal Fourier transform, which gives a
real-valued volume.

\setcounter{equation}{0}
\section{Mathematical formulation of Evo-MSSA}
\label{sec:appendix_evomssa}

This appendix gives the detailed form of the three modules summarized in
Algorithm~\ref{alg:evomssa}. All operations below are performed on one temporal
frequency slice. The noisy observed slice $Y_k$ and mask $M$ remain fixed over
the 30 reconstruction iterations, whereas $U_k^{(j)}$ is the current iterate.

\paragraph{Local Hankel embedding} For a slice $U_k^{(j)}$, the operator
$\mathcal H$ extracts all overlapping $6\times6$ spatial patches, vectorizes
them, and places them as columns of
\begin{equation}
	B_k^{(j)}
	=
	\mathcal H(U_k^{(j)})
	=
	\left[b_1,\ldots,b_{n_p}\right],
	\qquad b_p\in\mathbb C^{36}.
	\label{eq:app_mssa_hankel}
\end{equation}
Here, $n_p$ is the number of overlapping patches.
The associated operator $\mathcal H^\dagger$ returns the columns to their
original patch locations and forms a weighted overlap-add reconstruction using
a separable Hanning taper on the $6\times6$ support, normalized pointwise by
the accumulated taper weights. Thus, $\mathcal H^\dagger$ denotes tapered
overlap-averaging reconstruction, not a Moore--Penrose pseudoinverse.

\paragraph{MP--GD singular-value shrinkage} Consider either the complete
Hankel matrix or one of its column groups, denoted by
$B_g\in\mathbb C^{m_g\times n_g}$. Its left Gram decomposition is
\begin{equation}
	B_gB_g^{\mathrm H}
	=
	Q_g\operatorname{diag}
	\left(s_{g,1}^2,\ldots,s_{g,r_g}^2\right)Q_g^{\mathrm H},
	\qquad
	s_{g,1}\geq\cdots\geq s_{g,r_g}\geq0,
	\label{eq:app_mssa_gram}
\end{equation}
where $(\cdot)^{\mathrm H}$ is the conjugate transpose and
$r_g=\min(m_g,n_g)$. Let $\mathcal T_g$ be the lower half of this singular
spectrum, and let
$\eta_g=\min(m_g,n_g)/\max(m_g,n_g)$ be the measured matrix aspect ratio. If
$q_g^{\mathrm{MP}}$ denotes the 25th percentile of the Marchenko--Pastur
eigenvalue law \citep{marchenko1967distribution}, the matrix noise scale is
\begin{equation}
	\widehat\sigma_{B,g}
	=
	\frac{\operatorname{median}(\mathcal T_g)}
	{\sqrt{q_g^{\mathrm{MP}}}}.
	\label{eq:app_mssa_noise_scale}
\end{equation}
The 25th percentile appears because the median of the lower half of the full
spectrum is its first quartile. If the plug-in estimate is unavailable, the
implementation uses the largest finite value among the tail median, its
MAD-based scale, and the numerical floor $\epsilon$.

Let $\omega_\star(\eta_g)$ be the closed-form Gavish--Donoho optimal
hard-threshold coefficient \citep{gavish2014optimal}. Evo-MSSA sets
$\theta_g=\omega_\star(\eta_g)\widehat\sigma_{B,g}$ and forms the quadratic
shrinkage weights
\begin{equation}
	w_{g,i}
	=
	\left[1-\left(\frac{\theta_g}{s_{g,i}+\epsilon}\right)^2\right]_+.
	\label{eq:app_mssa_weights}
\end{equation}
Let $r_g^\star=\min(16,r_g)$ be the inherited rank limit after accounting for
the available matrix rank, and let $Q_{g,r_g^\star}$ contain the corresponding
leading left singular vectors. The MP--GD projection is
\begin{equation}
	\operatorname{MPGD}(B_g)
	=
	Q_{g,r_g^\star}\,
	\operatorname{diag}(w_{g,1},\ldots,w_{g,r_g^\star})\,
	Q_{g,r_g^\star}^{\mathrm H}B_g,
	\label{eq:app_mssa_mpgd}
\end{equation}
with components beyond the inherited rank limit assigned zero weight. In the
implementation, trailing zero-weight components are omitted. If all weights
vanish, the projected group is zero, after which overlap averaging and the soft
update proceed unchanged. The MP law therefore calibrates the noise scale, the
Gavish--Donoho form sets the threshold location, and
Eq.~\eqref{eq:app_mssa_weights} performs the actual shrinkage.
These quantities are used only as data-dependent calibration scales; no
optimality is claimed for the resulting quadratic shrinkage on the correlated
Hankel matrices considered here.

\paragraph{Reliability-grouped projection} No previous reliability field is
available at $j=0$, so Eq.~\eqref{eq:app_mssa_mpgd} is applied to the complete
matrix $B_k^{(0)}$. For $j\geq1$, let $\mathcal W_p$ be the spatial support of
column $b_p$. Its mean reliability and the resulting two groups are
\begin{equation}
	\begin{aligned}
		\xi_p^{(j-1)}
		&=
		\frac{1}{|\mathcal W_p|}
		\sum_{u\in\mathcal W_p}\Gamma_k^{(j-1)}(u),
		&
		\bar\xi^{(j-1)}
		&=
		\frac{1}{n_p}\sum_{p=1}^{n_p}\xi_p^{(j-1)},\\
		\mathcal I_-^{(j)}
		&=
		\{p:\xi_p^{(j-1)}<\bar\xi^{(j-1)}\},
		&
		\mathcal I_+^{(j)}
		&=
		\{p:\xi_p^{(j-1)}\geq\bar\xi^{(j-1)}\}.
	\end{aligned}
	\label{eq:app_mssa_groups}
\end{equation}
Each group is projected using its own spectrum and aspect ratio, and its
columns are then returned to their original positions:
\begin{equation}
	\widetilde B_{k,:,\,\mathcal I_\pm^{(j)}}^{(j)}
	=
	\operatorname{MPGD}
	\left(B_{k,:,\,\mathcal I_\pm^{(j)}}^{(j)}\right).
	\label{eq:app_mssa_group_projection}
\end{equation}
This operation is denoted by $\operatorname{GroupMPGD}$ in
Algorithm~\ref{alg:evomssa}.

\paragraph{Residual-based reliability} Let
$E_k^{(j)}=Y_k-L_k^{(j)}$ be the prediction residual, evaluated only where
$M=1$. Let $K$ be the
$6\times6$ all-ones support kernel and $*$ denote reflected-boundary spatial
convolution. The local observed mass, residual power, and observed power are
\begin{equation}
	m_K=K*M,\qquad
	P_E=K*\left(M\odot|E_k^{(j)}|^2\right),\qquad
	P_Y=K*\left(M\odot|Y_k|^2\right).
	\label{eq:app_mssa_local_power}
\end{equation}
The global MP noise scale from the first projection is retained as a
per-frequency anchor. With
$d_M=(n_xn_y)^{-1}\sum_u M(u)$ denoting the sampling density, the corresponding
noise power on observed samples is
\begin{equation}
	\widehat v_k
	=
	\frac{(\widehat\sigma_{B,k}^{(0)})^2}{n_p\,d_M+\epsilon}.
	\label{eq:app_mssa_noise_anchor}
\end{equation}
Two bounded local signal fractions are then
\begin{equation}
	a_{\mathrm{obs}}
	=
	\operatorname{clip}\left(
	1-\frac{P_E}{P_Y+\epsilon},0,1\right),
	\qquad
	a_{\mathrm{noise}}
	=
	\operatorname{clip}\left(
	1-\frac{\widehat v_k}
	{P_E/(m_K+\epsilon)+\epsilon},0,1\right).
	\label{eq:app_mssa_local_fractions}
\end{equation}
Writing $\pi_j=j/(J-1)$ for iteration progress, the raw local reliability and
its overlapping-support average are
\begin{equation}
	a_{\mathrm{raw}}^{(j)}
	=
	(1-\pi_j)a_{\mathrm{obs}}+\pi_j a_{\mathrm{noise}},
	\qquad
	a_{\mathrm{sp}}^{(j)}
	=
	\frac{K*(M\odot a_{\mathrm{raw}}^{(j)})}{m_K+\epsilon}.
	\label{eq:app_mssa_spatial_consensus}
\end{equation}

\paragraph{Cross-iteration consensus} Let the observed-sample mean and
centered spatial deviation be
\begin{equation}
	\mu_j
	=
	\frac{\sum_u M(u)a_{\mathrm{sp}}^{(j)}(u)}
	{\sum_u M(u)+\epsilon},
	\qquad
	D_j=a_{\mathrm{sp}}^{(j)}-\mu_j.
\end{equation}
Evo-MSSA forms a centered Ces\`aro average across iterations,
\begin{equation}
	\overline D_j
	=
	\begin{cases}
		D_0, & j=0,\\
		\overline D_{j-1}
		+\dfrac{D_j-\overline D_{j-1}}{j+1}, & j\geq1,
	\end{cases}
	\qquad
	\Gamma_{\mathrm{loc}}^{(j)}
	=
	\operatorname{clip}(\mu_j+\overline D_j,0,1).
	\label{eq:app_mssa_cross_iteration}
\end{equation}
This average reduces iteration-to-iteration fluctuations without introducing
a forgetting factor.

For robustness to complex residual outliers, let
$\mathcal E_k^{(j)}=\{E_k^{(j)}(u):M(u)=1\}$, $n_E=|\mathcal E_k^{(j)}|$, and
$c_E=\operatorname{median}(\Re\mathcal E_k^{(j)})
+\mathrm i\,\operatorname{median}(\Im\mathcal E_k^{(j)})$. For $n_E>0$, define
$v_{\mathrm{mean}}=n_E^{-1}\sum_{e\in\mathcal E_k^{(j)}}|e-c_E|^2$ and
\begin{equation}
	\begin{aligned}
		v_{\mathrm{rob}}
		&=
		\frac{\operatorname{median}
		\left(|\mathcal E_k^{(j)}-c_E|^2\right)}{\log 2},
		&
		\widetilde\pi_j
		&=
		\pi_j+\frac{1-\pi_j}{n_E},\\
		\Gamma_{\mathrm{rob}}^{(j)}
		&=
		(1-\widetilde\pi_j)
		+\widetilde\pi_j\,
		\operatorname{clip}\left[
		1-\left(1-\frac{1}{n_E}\right)
		\frac{v_{\mathrm{rob}}}{v_{\mathrm{mean}}+\epsilon},
		0,1\right].
	\end{aligned}
	\label{eq:app_mssa_robust_trust}
\end{equation}
Both $\pi_j$ and its finite-sample correction $\widetilde\pi_j$ gradually move
the update from conservative retention of observed samples to residual- and
noise-calibrated weighting.

The MP-anchored signal fraction and the final map used by
$\operatorname{ConsensusTrust}$ are
\begin{equation}
	\begin{aligned}
		\zeta_k
		&=
		\operatorname{clip}\left(
		1-
		\frac{\widehat v_k\sum_u M(u)}
		{\sum_u M(u)|Y_k(u)|^2+\epsilon},
		0,1\right),\\
		\Gamma_k^{(j)}
		&=
		(1-\zeta_k)\Gamma_{\mathrm{loc}}^{(j)}
		+\zeta_k\Gamma_{\mathrm{rob}}^{(j)}.
	\end{aligned}
	\label{eq:app_mssa_final_trust}
\end{equation}

\paragraph{Soft update and fallback rules} Overlap averaging gives
$L_k^{(j)}=\mathcal H^\dagger(\widetilde B_k^{(j)})$, and the next iterate is
\begin{equation}
	U_k^{(j+1)}
	=
	M\odot
	\left[
	\Gamma_k^{(j)}\odot Y_k
	+
	(1-\Gamma_k^{(j)})\odot L_k^{(j)}
	\right]
	+
	(1-M)\odot L_k^{(j)}.
	\label{eq:app_mssa_soft_update}
\end{equation}
Thus, large reliability values retain more of the noisy observation, whereas
small values place more weight on the low-rank prediction. The observed traces
are never hard-restored.

The first iteration, a malformed reliability field, nonfinite patch scores, a
group with fewer columns than Hankel rows, or a failed group eigendecomposition
triggers the global MP--GD projection. Invalid spectral estimates use the
finite tail-median/MAD fallback. If the observed-residual set is empty, the
slice-level reliability falls back to the observed signal fraction; an invalid
reconstructed frequency slice falls back to the observed slice. The final
operator uses all 30 reconstruction iterations. After all positive
frequency slices are processed, the inverse real Fourier transform produces a
real-valued reconstructed volume.

%% file: bibliography.bib
@article{gao2010irregular,
	title={Irregular seismic data reconstruction based on exponential threshold model of {POCS} method},
	author={Gao, Jian-Jun and Chen, Xiao-Hong and Li, Jing-Ye and Liu, Guo-Chang and Ma, Jian},
	journal={Applied Geophysics},
	volume={7},
	number={3},
	pages={229--238},
	year={2010},
	publisher={Springer}
}

@article{wang2014dreamlet,
	title={Dreamlet-based interpolation using {POCS} method},
	author={Wang, Benfeng and Wu, Ru-Shan and Geng, Yu and Chen, Xiaohong},
	journal={Journal of Applied Geophysics},
	volume={109},
	pages={256--265},
	year={2014},
	publisher={Elsevier}
}

@article{zhang20153d,
	title={{3D} seismic data reconstruction based on complex-valued curvelet transform in frequency domain},
	author={Zhang, Hua and Chen, Xiaohong and Li, Hongxing},
	journal={Journal of Applied Geophysics},
	volume={113},
	pages={64--73},
	year={2015},
	publisher={Elsevier}
}

@article{dong2025robust,
	title={Robust reconstruction of non-uniformly sampled {3D} seismic data with outliers},
	author={Dong, Lieqian and Zhang, Mugang and Xu, Yingjie and Wang, Changhui and Yu, Siwei and Zhang, Yingming},
	journal={Journal of Geophysics and Engineering},
	volume={22},
	number={2},
	pages={560--573},
	year={2025},
	publisher={Oxford University Press}
}

@article{zhang2016multi,
	title={Multi-step damped multichannel singular spectrum analysis for simultaneous reconstruction and denoising of {3D} seismic data},
	author={Zhang, Dong and Chen, Yangkang and Huang, Weilin and Gan, Shuwei},
	journal={Journal of Geophysics and Engineering},
	volume={13},
	number={5},
	pages={704--721},
	year={2016},
	publisher={Oxford University Press}
}

@article{bayati20233,
	title={{3-D} data interpolation and denoising by an adaptive weighting rank-reduction method using multichannel singular spectrum analysis algorithm},
	author={Bayati, Farzaneh and Trad, Daniel},
	journal={Sensors},
	volume={23},
	number={2},
	pages={577},
	year={2023},
	publisher={MDPI}
}

@article{huang2016damped,
	title={Damped multichannel singular spectrum analysis for {3D} random noise attenuation},
	author={Huang, Weilin and Wang, Runqiu and Chen, Yangkang and Li, Huijian and Gan, Shuwei},
	journal={Geophysics},
	volume={81},
	number={4},
	pages={V261--V270},
	year={2016},
	publisher={Society of Exploration Geophysicists}
}

@article{spitz1991seismic,
	title={Seismic trace interpolation in the {F-X} domain},
	author={Spitz, Simon},
	journal={Geophysics},
	volume={56},
	number={6},
	pages={785--794},
	year={1991},
	publisher={Society of Exploration Geophysicists},
	doi={10.1190/1.1443096}
}

@article{abma20063d,
	title={{3D} interpolation of irregular data with a {POCS} algorithm},
	author={Abma, Ray and Kabir, Nurul},
	journal={Geophysics},
	volume={71},
	number={6},
	pages={E91--E97},
	year={2006},
	publisher={Society of Exploration Geophysicists},
	doi={10.1190/1.2356088}
}

@article{oropeza2011simultaneous,
	title={Simultaneous seismic data denoising and reconstruction via multichannel singular spectrum analysis},
	author={Oropeza, Vicente and Sacchi, Mauricio D.},
	journal={Geophysics},
	volume={76},
	number={3},
	pages={V25--V32},
	year={2011},
	publisher={Society of Exploration Geophysicists},
	doi={10.1190/1.3552706}
}

@article{yu2026evolutionary,
	title={Evolutionary Ensemble of Agents},
	author={Yu, Zongmin and Yang, Liu},
	journal={arXiv preprint arXiv:2605.09018},
	year={2026},
	eprint={2605.09018},
	archivePrefix={arXiv},
	primaryClass={cs.NE},
	doi={10.48550/arXiv.2605.09018},
	url={https://arxiv.org/abs/2605.09018}
}

@article{wu2018adaptive,
  title={Adaptive rank-reduction method for seismic data reconstruction},
  author={Wu, Juan and Bai, Min},
  journal={Journal of Geophysics and Engineering},
  volume={15},
  number={4},
  pages={1688--1703},
  year={2018},
  doi={10.1093/jge/aabc74}
}

@article{marchenko1967distribution,
	title={Distribution of eigenvalues for some sets of random matrices},
	author={Marchenko, Vladimir A. and Pastur, Leonid A.},
	journal={Mathematics of the USSR-Sbornik},
	volume={1},
	number={4},
	pages={457--483},
	year={1967},
	doi={10.1070/SM1967v001n04ABEH001994}
}

@article{gavish2014optimal,
	title={The optimal hard threshold for singular values is $4/\sqrt{3}$},
	author={Gavish, Matan and Donoho, David L.},
	journal={IEEE Transactions on Information Theory},
	volume={60},
	number={8},
	pages={5040--5053},
	year={2014},
	doi={10.1109/TIT.2014.2323359}
}

@article{chen2020odrr,
	title={Five-dimensional seismic data reconstruction using the optimally damped rank-reduction method},
	author={Chen, Yangkang and Bai, Min and Guan, Zhe and Zhang, Qingchen and Zhang, Mi and Wang, Hang},
	journal={Geophysical Journal International},
	volume={222},
	number={3},
	pages={1824--1845},
	year={2020},
	doi={10.1093/gji/ggaa190}
}

@article{chen2023drr,
	title={{DRR}: An open-source multi-platform package for the damped rank-reduction method and its applications in seismology},
	author={Chen, Yangkang and Huang, Weilin and Yang, Liuqing and Obou{\'e}, Yapo Abol{\'e} Serge Innocent and Saad, Omar M. and Chen, Yunfeng},
	journal={Computers \& Geosciences},
	volume={180},
	pages={105440},
	year={2023},
	doi={10.1016/j.cageo.2023.105440}
}

@article{trad2009five,
	title={Five-dimensional interpolation: Recovering from acquisition constraints},
	author={Trad, Daniel},
	journal={Geophysics},
	volume={74},
	number={6},
	pages={V123--V132},
	year={2009},
	publisher={Society of Exploration Geophysicists}
}

@article{herrmann2008non,
	title={Non-parametric seismic data recovery with curvelet frames},
	author={Herrmann, Felix J. and Hennenfent, Gilles},
	journal={Geophysical Journal International},
	volume={173},
	number={1},
	pages={233--248},
	year={2008},
	publisher={Oxford University Press},
	doi={10.1111/j.1365-246X.2007.03698.x}
}

@article{gao2013convergence,
	title={Convergence improvement and noise attenuation considerations for beyond alias projection onto convex sets reconstruction},
	author={Gao, Jianjun and Stanton, Aaron and Naghizadeh, Mostafa and Sacchi, Mauricio D. and Chen, Xiaohong},
	journal={Geophysical Prospecting},
	volume={61},
	number={S1},
	pages={138--151},
	year={2013},
	publisher={Wiley},
	doi={10.1111/j.1365-2478.2012.01103.x}
}

@article{naghizadeh2007multistep,
	title={Multistep autoregressive reconstruction of seismic records},
	author={Naghizadeh, Mostafa and Sacchi, Mauricio D},
	journal={Geophysics},
	volume={72},
	number={6},
	pages={V111--V118},
	year={2007},
	publisher={Society of Exploration Geophysicists}
}

@article{xu2024dealiased,
	title={Dealiased seismic data interpolation by dynamic matching},
	author={Xu, Yingjie and Yu, Siwei and Dong, Lieqian and Ma, Jianwei},
	journal={Geophysics},
	volume={89},
	number={5},
	pages={V361--V376},
	year={2024},
	publisher={Society of Exploration Geophysicists},
	doi={10.1190/geo2023-0249.1}
}

@article{yu2019deep,
	author  = {Yu, Siwei and Ma, Jianwei and Wang, Wenlong},
	title   = {Deep learning for denoising},
	journal = {Geophysics},
	year    = {2019},
	volume  = {84},
	number  = {6},
	pages   = {V333--V350},
	doi     = {10.1190/geo2018-0668.1}
}

@article{wang2019deep,
	author  = {Wang, Benfeng and Zhang, Ning and Lu, Wenkai and Wang, Jialin},
	title   = {Deep-learning-based seismic data interpolation: A preliminary result},
	journal = {Geophysics},
	year    = {2019},
	volume  = {84},
	number  = {1},
	pages   = {V11--V20},
	doi     = {10.1190/geo2017-0495.1}
}

@article{kaur2021seismic,
	author  = {Kaur, Harpreet and Pham, Nam and Fomel, Sergey},
	title   = {Seismic data interpolation using deep learning with generative adversarial networks},
	journal = {Geophysical Prospecting},
	year    = {2021},
	volume  = {69},
	number  = {2},
	pages   = {307--326},
	doi     = {10.1111/1365-2478.13055}
}

@article{cheng2025multitask,
	author  = {Cheng, Ming and Lin, Jun and Dong, Xintong and Zhong, Tie},
	title   = {A multitask deep-learning model for the denoising, interpolation, and wavefield separation of distributed acoustic sensing-vertical seismic profiling data},
	journal = {Geophysics},
	year    = {2025},
	volume  = {90},
	number  = {6},
	pages   = {V559--V568},
	doi     = {10.1190/geo2024-0531.1}
}

@article{liu2021deep,
	title={Deep-seismic-prior-based reconstruction of seismic data using convolutional neural networks},
	author={Liu, Qun and Fu, Lihua and Zhang, Meng},
	journal={Geophysics},
	volume={86},
	number={2},
	pages={V131--V142},
	year={2021},
	publisher={Society of Exploration Geophysicists},
	doi={10.1190/geo2019-0570.1}
}

@article{chen2024combining,
	title={Combining unsupervised deep learning and {Monte Carlo} dropout for seismic data reconstruction and its uncertainty quantification},
	author={Chen, Gui and Liu, Yang},
	journal={Geophysics},
	volume={89},
	number={1},
	pages={WA53--WA65},
	year={2024},
	publisher={Society of Exploration Geophysicists},
	doi={10.1190/geo2022-0632.1}
}

@article{xu2026unsupervised,
	author  = {Xu, Ying-Jie and Wang, Wen-Chuang and Yu, Si-Wei and Ma, Jian-Wei},
	title   = {Unsupervised anti-aliasing interpolation of regularly sampled seismic data via soft dynamic time warping divergence distance},
	journal = {Petroleum Science},
	year    = {2026},
	doi     = {10.1016/j.petsci.2026.06.031},
	note    = {In press}
}

@article{abedi2022multidirectional,
	author  = {Abedi, Mohammad Mahdi and Pardo, David},
	title   = {A multidirectional deep neural network for self-supervised reconstruction of seismic data},
	journal = {IEEE Transactions on Geoscience and Remote Sensing},
	year    = {2022},
	volume  = {60},
	pages   = {1--9},
	doi     = {10.1109/TGRS.2022.3227212}
}

@article{meng2022self,
	author  = {Meng, Fanlei and Fan, QinYin and Li, Yue},
	title   = {Self-supervised learning for seismic data reconstruction and denoising},
	journal = {IEEE Geoscience and Remote Sensing Letters},
	year    = {2022},
	volume  = {19},
	pages   = {1--5},
	doi     = {10.1109/LGRS.2021.3068132}
}

@article{sheng2025seismic,
	author  = {Sheng, Hanlin and Wu, Xinming and Si, Xu and Li, Jintao and Zhang, Sibo and Duan, Xudong},
	title   = {Seismic foundation model: A next generation deep-learning model in geophysics},
	journal = {Geophysics},
	year    = {2025},
	volume  = {90},
	number  = {2},
	pages   = {IM59--IM79},
	doi     = {10.1190/geo2024-0262.1}
}

@article{cheng2025generative,
	author  = {Cheng, Shijun and Harsuko, Randy and Alkhalifah, Tariq},
	title   = {A generative foundation model for an all-in-one seismic processing framework},
	journal = {Surveys in Geophysics},
	year    = {2025},
	volume  = {46},
	number  = {6},
	pages   = {1173--1215},
	doi     = {10.1007/s10712-025-09912-9}
}

@article{romeraparedes2024mathematical,
title={Mathematical discoveries from program search with large language models},
author={Romera-Paredes, Bernardino and Barekatain, Mohammadamin and Novikov, Alexander and Balog, Matej and Kumar, M. Pawan and Dupont, Emilien and Ruiz, Francisco J. R. and Ellenberg, Jordan S. and Wang, Pengming and Fawzi, Omar and Kohli, Pushmeet and Fawzi, Alhussein},
journal={Nature},
volume={625},
number={7995},
pages={468--475},
year={2024},
publisher={Nature Publishing Group},
doi={10.1038/s41586-023-06924-6}
}

@inproceedings{liu2024evolution,
title={Evolution of Heuristics: Towards Efficient Automatic Algorithm Design Using Large Language Model},
author={Liu, Fei and Tong, Xialiang and Yuan, Mingxuan and Lin, Xi and Luo, Fu and Wang, Zhenkun and Lu, Zhichao and Zhang, Qingfu},
booktitle={Proceedings of the 41st International Conference on Machine Learning},
series={Proceedings of Machine Learning Research},
volume={235},
pages={32201--32223},
year={2024},
publisher={PMLR},
url={https://proceedings.mlr.press/v235/liu24bs.html}
}

@inproceedings{ye2024reevo,
title={{ReEvo}: Large Language Models as Hyper-Heuristics with Reflective Evolution},
author={Ye, Haoran and Wang, Jiarui and Cao, Zhiguang and Berto, Federico and Hua, Chuanbo and Kim, Haeyeon and Park, Jinkyoo and Song, Guojie},
booktitle={Advances in Neural Information Processing Systems},
volume={37},
pages={43571--43608},
year={2024},
publisher={Curran Associates, Inc.},
url={https://proceedings.neurips.cc/paper_files/paper/2024/hash/4ced59d480e07d290b6f29fc8798f195-Abstract-Conference.html}
}

@article{novikov2025alphaevolve,
title={{AlphaEvolve}: A coding agent for scientific and algorithmic discovery},
author={Novikov, Alexander and V{\~u}, Ng{\^a}n and Eisenberger, Marvin and Dupont, Emilien and Huang, Po-Sen and Wagner, Adam Zsolt and Shirobokov, Sergey and Kozlovskii, Borislav and Ruiz, Francisco JR and Mehrabian, Abbas and others},
journal={arXiv preprint arXiv:2506.13131},
year={2025},
doi={10.48550/arXiv.2506.13131}
}

@inproceedings{ma2024eureka,
author    = {Ma, Yecheng Jason and Liang, William and Wang, Guanzhi
and Huang, De-An and Bastani, Osbert and Jayaraman, Dinesh
and Zhu, Yuke and Fan, Linxi and Anandkumar, Anima},
title     = {{Eureka}: Human-Level Reward Design via Coding Large
Language Models},
booktitle = {The Twelfth International Conference on Learning
Representations},
year      = {2024},
url       = {https://openreview.net/forum?id=IEduRUO55F}
}

@article{sun2026self,
  title={Self-Evolving Scientific Agent Discovers Generalizable Physically-Reasoned Fluid Control},
  author={Sun, Boai and Guo, Wenjin and Yu, Zongmin and Yang, Liu},
  journal={arXiv preprint arXiv:2606.08405},
  year={2026}
}

@article{yang2026harness,
  title={Harness In-Context Operator Learning with Chain of Operators},
  author={Yang, Minghui and Guo, Ling and Yang, Liu},
  journal={arXiv preprint arXiv:2606.12318},
  year={2026}
}

@article{yu2026agentic,
  title={Agentic Symbolic Search: Characterizing PDEs Beyond Hand-crafted Expressions, Meshes, and Neural Networks},
  author={Yu, Zongmin and Yang, Liu},
  journal={arXiv preprint arXiv:2606.20467},
  year={2026}
}
